\documentclass[twocolumn,reprint, aps,pra,longbibliography]{revtex4-2}
\usepackage{amsmath,latexsym,amssymb}
\usepackage[]{mathrsfs}
\usepackage[section]{placeins}
\usepackage{placeins}
\usepackage{graphicx}
\usepackage{dcolumn}
\usepackage{bm}
\usepackage[]{comment}
\usepackage[mathscr]{euscript}
\usepackage{hyperref} 
\newcommand{\vect}[1]
{\vec{\boldsymbol{#1}}}

\newcommand{\stat}[1]
{\left|#1\right\rangle}

\newcommand{\bra}[1]
{\left\langle#1\right|}

\newcommand{\ip}[2]
{\langle{#1}\vert{#2}\rangle}
\newcommand{\abs}[1]
{\left\vert#1\right\vert}

\begin{document}
	\title{Scouring Parrondo’s Paradox in Discrete-Time Quantum Walks}
	
	\author{Gururaj Kadiri}
	\email{gururaj@igcar.gov.in}
	\affiliation{Materials Science Group, Indira Gandhi Centre for Atomic Research, Kalpakkam, Tamilnadu, 603102, India.} 
	\begin{abstract}
		We propose a quantum game based on coin-based quantum walks. Given a quantum walk and a Hermitian operator on the coin-position composite space, winning this game involves choosing an initial coin state such that the given quantum walk leads to a composite state in which the expectation value of the given Hermitian operator is greater than a certain value. Parrondo’s paradox is a phenomenon where a combination of losing strategies becomes a winning strategy. We give a deterministic scheme for identifying Parrondo's paradox in our game, in the sense that, given a collection of distinct quantum steps, we identify initial coin states which happen to be losing states for all quantum walks comprising solely of these steps individually, but turn out to be winning states for a quantum walk comprising of all the given steps taken in a sequence. Unlike traditional quantum steps that allow for equal magnitude forward and backward strides based on the outcome of the coin-toss, the steps of the quantum walks employed here, though still contingent upon coin-toss, permit the strides to be of unequal magnitude, and not necessarily in opposite directions. We believe the results presented here will contribute to a deeper understanding of evolution of expectation values of observables in quantum walks, and facilitate the development of novel quantum algorithms.
	\end{abstract}
	\maketitle
	\section{Introduction}
Game theory is an interdisciplinary framework that analyzes strategic interactions among rational decision-makers. This mathematical discipline provides a systematic approach to understanding decision-making in competitive situations and provides a robust framework for optimizing individual actions \cite{owen2013game,maschler2020game,kolokoltsov2020understanding}. Parrondo's paradox is a seemingly counter-intuitive phenomenon presented in this theory, wherein two games, both unfavorable when played individually, can become favorable when played together in a particular sequence \cite{parrondo2000new,harmer1999losing,lai2020parrondo,abbott2010asymmetry}. 
This captivating paradox has found applications in fields as diverse as dynamical systems \cite{canovas2013revisiting,danca2014generalized,arena2003game}, control theory \cite{allison2001control}, sociology \cite{lai2020social,lai2024parrondo}, biology \cite{cheong2019paradoxical,cheong2020relieving}, financial markets \cite{bassi2011parrondos}, to name a few. To elucidate the concept of this paradox in a straightforward manner, let’s examine the scenario of utilizing an escalator to ascend to the top floor, starting at the middle. Assume we have at our disposal two escalators A and B, which are such that escalator A always moves down one step at every unit of time, and escalator B moves two steps up and three steps down in alternate units of time. Evidently, both the escalators are unfavorable in the long run, in the sense that neither of them can lead us to the top of the floor in a finite amount of time. However, consider the choice of alternating the use of two escalators: that is, using the escalator B during the times in which it is moving up, and using escalator A during the time in which the escalator B is moving down. This switching strategy will lead to a net movement of one step forward in two units of time, eventually taking us to the top floor. This example provides a simple illustration of the Parrondo's paradox which can be termed as ``winning by alternating between two losing strategies''. \par
Quantum games are games where players have access to quantum resources like entanglement and superposition, which influence their strategies and payoffs. These games often played using qubits as players and quantum operations as moves.  In this context, quantum game theory refers to the conceptual tool for exploring how the availability of quantum resources affect strategic decision-making and cooperative behavior \cite{flitney2002introduction,khan2018quantum,eisert1999quantum,du2002experimental}. Parrondo's paradox, already challenging classical rationality by demonstrating unexpected gains from seemingly losing strategies, takes on a new dimension within the quantum game regime, since these games enable interference between probability amplitudes across different game paths, as against classical Parrondo games, where probabilities simply add up. \par
Quantum walks refer to the theoretical framework for describing the movement of quantum particles through a discrete or continuous space based on the principles of quantum mechanics \cite{portugal2013quantum}. These walks are the quantum-mechanical counterpart to the ubiquitous classical walks, and are one of the most-suited platforms for demonstrating the essential quantum phenomena like superposition, post-measurement collapse etc \cite{mackay2002quantum,VenegasAndraca2012}. Quantum walks have been shown to be a paradigm for quantum computation \cite{lovett2010universal,singh2021universal,asaka2023two}, and they have proven to be a powerful tool in the quantum algorithm designer’s toolkit, enabling the creation of algorithms that can outperform classical ones for certain class of problems\cite{kendon2006random,childs2003exponential,apers2022quadratic,lovett2019quantum}. 
These walks have been experimentally demonstrated in different hardware platforms like circuit qed, linear optics, trapped ions etc \cite{zhou2019protocol,su2019experimental,giordani2019experimental,matjeschk2012experimental}. \par
This paper explores the conditions necessary for the manifestation of Parrondo's paradox in some single-player games designed within the context of a specific variant of quantum walks, called the discrete-time quantum walks (DTQWs). These walks play out on the tensor product of two vector spaces: a two-dimensional vector space called the coin space, and a $k$-dimensional space, called the position space. In a typical DTQW, the walker starts from the position zero in a definite coin state. At every step, she proceeds forward or backwards by one unit conditional to the outcome of a coin-toss operator, which can be modeled as an unitary operator on the coin-space. Since the coin-toss operation is not followed by a measurement, in one quantum step the quantum state evolves in both forward and backward directions coherently on the position space. A quantum walk corresponds to an application of multiple number of such quantum steps. The outcome of such a quantum walk is an entangled quantum state, with the probability amplitude spread over multiple positions. \par
One of the earliest references to the Parrondo's paradox in DTQWs is in ref. \cite{flitney2004quantum}, where a model of DTQW that depends on the outcomes of previous coin tosses is shown to display the Parrondo's paradox. This paradox was demonstrated in discrete walks employing three-sided and four-sided, time-dependent coins and also aperodic sequence of coins \cite{walczak2022parrondothreecoins,lai2020parrondofoursided,pires2020parrondo,walczak2021parrondo}. 	The possibility of realizing this paradox in the Bose-Einstein condensate is examined in \cite{trautmann2022parrondo}. The relation of this paradox to the coin-position entanglement in DTQWs has been studied in \cite{panda2022generating} and \cite{fang2023maximal}, and the ability of noise to induce Parrondo's paradox in DTQWs was studied in ref. \cite{walczak2023noise}. 
A proposal to deploy a quantum coin toss for encryption has been proposed in \cite{lai2021chaotic}, and an experimental demonstration of this paradox in DTQWs within the quantum optics scenario is presented in ref \cite{jan2020experimental}.  
\par
Before discussing the contribution of this paper, it is important to briefly elucidate the general scheme that a majority of these papers have adopted for demonstrating this paradox. Owing to the translationally invariant action of a quantum walk, to every state on the coin-space one can associate a unique state on the composite space. This state is the outcome of initiating the quantum walk from the zero-position in that coin state. In the context of quantum games on DTQWs, a coin state was termed a winning state if in its associated composite state the probability of finding the walker at positive positions exceeds that at negative positions. The investigation into Parrondo’s paradox in discrete-time quantum walks commences with two distinct quantum steps. From these steps, two quantum walks having identical number of steps are constructed, each consisting of one kind of steps. Subsequently, a third quantum walk is built, having the same number of steps as the former two, but built out of both the steps applied in a specific sequence. Evidently, a coin state could be either a winning or losing state for each of the three walks separately. Now, a coin-state is said to display the ``Parrondo effect'' or ``Parrondo paradox'' if it is a losing state for the former two walks, but happens to be a winning state for the third walk. This is in line with the standard interpretation of Parrondo's paradox, since for such coin states the former two walks lead to a loss, but the third walk, even though comprised of same steps as the former two, leads to a winning. It is, as it were, these coin states employ two losing quantum steps in a specific sequence and win the game. 
\par In most of the papers demonstrating the Parrondo's paradox in discrete time quantum walks, the paradox is illustrated for some choice of quantum step pairs and for some initial coin states, see for instance ref. \cite{lai2020parrondoPRE,jan2023territories,chandrashekar2011parrondo,rajendran2018implementing}. To the best of our knowledge, there isn't any literature on how to identifying all the coin states that manifest the Parrondo paradox for a given pair of arbitrary quantum steps. In this backdrop, the contribution of this paper is the following: given a pair of quantum steps, we give a specific scheme for identifying all the coin-states which demonstrate the Parrondo's paradox. Actually, here we go much further, as listed below:
\begin{enumerate}
	\item We employ a more generalized definition of quantum steps and quantum walks. 
	\item We propose a generalization of winning quantum games, in the context of discrete time quantum walks.
	\item We define Parrondo's paradox in the context of multiple such quantum steps rather than two as done currently in the literature.
\end{enumerate}
While novel, it will be argued subsequently that we have expanded these definitions seamlessly from their conventional counterparts, staying true to the original. \par 
In relation to the first point, in this paper we employ quantum steps which could be biased in the forward or backward directions. These quantum steps are in-spirit similar to these employed in the split-step quantum walks \cite{Kitagawa2010,narimatsu2021unitary,matsuzawa2020index}. Like the standard quantum steps, our quantum steps also act in a translationally invariant manner conditional to a coin-toss, except that their stride-lengths in forward and backward directions not identical in magnitude. Indeed, the definition of quantum walks employed here encompasses even such steps wherein the walker moves along the same direction (forward or backward), but with different stride lengths depending on the output of the coin toss. 
\par With respect to the second point of defining a new winning criterion, we do it as follows. We begin by choosing an observable on the coin-position composite space. Given this observable as a Hermitian operator, to every coin we associate a real number called its ``payoff'', which is defined as the expectation value of the Hermitian operator in its associated composite state. Evidently, the payoff of a coin-state depends both on the observable and the quantum walk.  Now, a coin-state is considered a winning state if its payoff is greater than a certain real number, which we call as the ``target payoff''. This definition of winning includes the traditional definition as well. This is due to the existence of an observable in the position space, the expected value of which equals the difference between the probabilities of occupying positive and negative positions in that state. Another example of an operator on the position space is the position operator, whose expectation value in a certain state gives the mean position. With this operator, and taking the target payoff as zero, the winning coin states will happen to be those which lead to a composite state where in the mean position is positive. \par
With respect to the third point, we shall demonstrate the Parrondo paradox over a collection of multiple number of quantum steps, rather than just two as is traditionally done. \par

The rest of the paper is organized as follows. In Section [\ref{sec:Thrtical_Background}] we set-up the required theoretical and mathematical tools to be used for the rest of the section. We describe here the conventional quantum steps and quantum walks, and the notion of translational invariance of quantum states. we then introduce the novel quantum steps that we have conceived, and discuss their action. We also compare the contrast the two types of quantum steps. In Section [\ref{sec:PP_In_QW}] we give mathematical definition of the quantum games, winning them and the notion of Parrondo's paradox in these games. Section [\ref{sec:PP_in_ArbWalks}] constitutes the central contribution of this paper. Here, given a collection of quantum steps, we give a deterministic scheme for identifying the Parrondo states for the collection of walks for a given Hermitian operator on the composite space. These quantum steps could be standard quantum steps or the novel quantum steps introduced in this paper. In Section [\ref{sec:Accentuate_PP}], we demonstrate a scheme for realizing Parrondo paradox explicitly using the biased quantum steps defined here.  In Section [\ref{sec:Numerical_Illustration}], we provide a few numerical illustrations of the proposals given in the earlier sections. We conclude the paper by summarizing the central results in Section [\ref{sec:Conclusions}].
	\section{Theoretical Background }
	\label{sec:Thrtical_Background}
	We shall indicate the coin and position spaces by the symbol $H_c$ and $H_p$ respectively, and the composite space by $H \equiv H_c\otimes H_p$.
	\subsection{Definition of the quantum states}
	An arbitrary pure state of the coin space $H_c$ is given by:
	\begin{equation}
		\stat{\boldsymbol{s}}=s_0\stat{\boldsymbol{0}}+s_1\stat{\boldsymbol{1}},
		\label{eq:coin_space_state}
	\end{equation}
where $s_0$ and $s_1$ are complex numbers $\abs{s_0}^2+\abs{s_1}^2=1$,  and $\{\stat{\boldsymbol{0}},\stat{\boldsymbol{1}}\}$ constitute an orthogonal basis of $H_c$. Given a pure state $\stat{\boldsymbol{s}}$ of the above form, we associate with it another pure state $\stat{\boldsymbol{s}_\perp}$, defined as:
\begin{equation}
		\stat{\boldsymbol{s}_\perp}=\bar{s_0}\stat{\boldsymbol{1}}-\bar{s_1}\stat{\boldsymbol{0}},
\label{eq:perp_state}
\end{equation}
where the overbar indicates complex conjugation, as its orthogonal state. A mixed state of the coin space can be written as the density matrix
\begin{equation}
\hat{\rho}_{r,\boldsymbol{s}}=r\stat{\boldsymbol{s}}\bra{\boldsymbol{s}}+(1-r)\stat{\boldsymbol{s}_\perp}\bra{\boldsymbol{s}_\perp}
\label{eq:mixed_st_defintion}
\end{equation}
where $r$ is a fraction, $0\le r \le 1$. 
Given a density matrix of the form Eq. (\ref{eq:mixed_st_defintion}), we associate a vector in 3D, called here as the ``qubit vector'', defined as 
\begin{equation}
\vect{S}(\hat{\rho}_{r,\boldsymbol{s}})=(\textbf{tr}(\hat{\rho}_{r,\boldsymbol{s}}\hat{\sigma}_x),\textbf{tr}(\hat{\rho}_{r,\boldsymbol{s}}\hat{\sigma}_y),\textbf{tr}(\hat{\rho}_{r,\boldsymbol{s}}\hat{\sigma}_z))
\end{equation}
where $\hat{\sigma_i}, \, i=x,y,z$ are the $2\times 2$ operators defined as:
\begin{equation}
\begin{aligned}
\hat{\sigma_x}&=\stat{\boldsymbol{0}}\bra{\boldsymbol{1}}+\stat{\boldsymbol{1}}\bra{\boldsymbol{0}},\\\hat{\sigma_y}&=i(\stat{\boldsymbol{1}}\bra{\boldsymbol{0}}-\stat{\boldsymbol{0}}\bra{\boldsymbol{1}}), \text{ and }\\\hat{\sigma_z}&=\stat{\boldsymbol{0}}\bra{\boldsymbol{0}}-\stat{\boldsymbol{1}}\bra{\boldsymbol{1}}.
\end{aligned}
\end{equation}
The definition of qubit vector in the case of a pure state gets simplified to:
\begin{equation}
	\vec{\boldsymbol{S}}(\stat{\boldsymbol{s}})=\left(\ip{\boldsymbol{s}}{\hat{\sigma_x}|\boldsymbol{s}},\ip{\boldsymbol{s}}{\hat{\sigma_y}|\boldsymbol{s}},\ip{\boldsymbol{s}}{\hat{\sigma_z}|\boldsymbol{s}}\right)
	\label{eq:stokes_vector_pure_states}
\end{equation}
When the quantum state is expressed without regard to the global phase, we represent the state by the symbol $\stat{\boldsymbol{\theta,\phi}}$:
\begin{equation}
	\stat{\boldsymbol{\theta,\phi}}=\cos\frac{\theta}{2}\stat{\boldsymbol{0}}+e^{i\phi}\sin\frac{\theta}{2}\stat{\boldsymbol{1}}
\end{equation}
For the position space, we assume a privileged orthonormal basis $\{\stat{m},m=-N,\cdots,N\}$, where $N$ is potentially infinite, and in numerical simulations here we take it to be larger than any position we access in a quantum walks. Localized pure-states in the coin-position composite space are of the form
	\begin{equation}
	\stat{\boldsymbol{s};m}=\stat{\boldsymbol{s}}\otimes \stat{m}
	\label{eq:product state}
		\end{equation}
A general composite pure-state is a superposition of these localized states as :
	\begin{equation}
		\stat{\boldsymbol{S}}\equiv\sum_{m=b}^{m=e}s_m\stat{\boldsymbol{s}_m;m}
		\label{eq:composite_state}
	\end{equation}
where $s_m$ are real numbers such that $\sum_{m}s_m^2=1$, and $b$ and $e\ge b$ are integers, indicating the span of positions involved in the superposition. The states $\stat{\boldsymbol{s}_m;m}$ are localized states of the form Eq. (\ref{eq:product state}). The sifted state and orthogonal states of $\stat{\boldsymbol{S}}$, indicated by $\stat{\boldsymbol{S}_{+d}}$ and $\stat{\boldsymbol{S}_\perp}$ respectively, are given by
\begin{equation}
	\begin{aligned}
		\stat{\boldsymbol{S}_{+d}}&=\sum_{m=b}^{m=e}s_m\stat{\boldsymbol{s}_m;m+d} \text{ and }\\
		\stat{\boldsymbol{S}_{\perp}}&=\sum_{m=b}^{m=e}s_m\stat{{(\boldsymbol{s}_m)}_\perp;-m}
	\end{aligned}
\end{equation}
It is easy to see that $\stat{\boldsymbol{S}_\perp}$ is orthogonal to the state $\stat{\boldsymbol{S}}$. 
We define a composite state $\stat{\boldsymbol{W}}$ of the form Eq. (\ref{eq:composite_state}) as being translationally invariant \cite{kadiri2023steered} if it satisfies the following orthogonality condition:
\begin{equation}
	\ip{\boldsymbol{W}}{\boldsymbol{W}_{+d}}=0, \, \forall d=1,\cdots,e-b.
	\label{eq:TI_definition}
\end{equation}
\par
A Hermitian operator $\hat{O}$ on the composite space $H_c\otimes H_p$ can be written as:
\begin{equation}
	\hat{O}=\sum_{i=1}^{2(2N+1)}\lambda_i\stat{\boldsymbol{U}_i}\bra{\boldsymbol{U}_i},
	\label{eq:operator_O}
\end{equation}
where $\lambda_i$ are real numbers, and $\{\stat{\boldsymbol{U}_i},i=1,\cdots,2(2N+1)\}$ is an orthogonal basis on the composite space, with each $\stat{\boldsymbol{U}_i}$ being of the form Eq. (\ref{eq:composite_state}). A density matrix is on the composite space is given by:
\begin{equation}
	\hat{\rho}=\sum_{i=1}^{2(2N+1)}r_i\stat{\boldsymbol{W}_i}\bra{\boldsymbol{W}_i}
\end{equation}
where $r_i$ are real numbers between $0$ and $1$, such that $\sum_{n}r_n=1$, and $\{\stat{\boldsymbol{W}_i},i=1,\cdots,2(2N+1)\}$ is some orthogonal basis on the composite space. 
\subsection{Conventional quantum steps and quantum walks}
We denote an arbitrary quantum step by the symbol  $\hat{\mathcal{T}}(\alpha,\beta,\gamma)$, where $\alpha$, $\beta$ and $\gamma$ are real numbers, with $0\le\alpha,\gamma<2\pi$ and $0\le\beta\le\frac{\pi}{2}$. This quantum step is defined in terms of two sub-steps as:
\begin{equation}
	\hat{\mathcal{T}}(\alpha,\beta,\gamma)=\hat{\mathcal{S}}(\hat{\mathit{c}}(\alpha,\beta,\gamma)\otimes \hat{I}_w),
	\label{eq:traditional_step}
\end{equation}
where $\hat{\mathcal{S}}$ and $\hat{\mathit{c}}$ are called the shift and coin-toss operator respectively, defined as:
\begin{equation}
	\begin{aligned}
\hat{\mathcal{S}}&=\sum_{n}\left(\stat{\boldsymbol{0};n-1}\bra{\boldsymbol{0};n}+\stat{\boldsymbol{1};n+1}\bra{\boldsymbol{1};n}\right),\\
\hat{\mathit{c}}(\alpha,\beta,\gamma)&=\left(\begin{array}{cc}
	e^{i\alpha}\cos\beta&  -e^{-i\gamma}\sin\beta\\
	e^{i\gamma}\sin\beta& e^{-i\alpha}\cos\beta
\end{array}\right).
\label{eq:traditional_shift}
\end{aligned}
\end{equation}
Here $\hat{\mathit{c}}(\alpha,\beta,\gamma)$ is an SU(2) operator \cite{chandrashekar2008optimizing} acting on the coin-space alone, while $\hat{\mathcal{S}}$ acts on position-space alone, but conditional to the coin state. A collection of such steps is defined as a quantum walk.
These quantum walks however cannot generate all possible translationally invariant quantum states from home-states. That is, not all translationally invariant states can be generated by employing such quantum steps alone. A simplest illustration is the two position state $\stat{\boldsymbol{S}}=s_m\stat{\boldsymbol{s}_m;m}+s_n\stat{\boldsymbol{s}_n;n}$ with $m\neq n$, the coin-state $\stat{\boldsymbol{s}_m}$ and $\stat{\boldsymbol{s}_n}$ being orthogonal: $\ip{\boldsymbol{s}_m}{\boldsymbol{s}_n}=0$, and $s_m$ and $s_n$ are non-zero complex numbers such that $s_m^2+s_n^2=1$. This state is a composite state spread over two positions $\stat{m}$ and $\stat{n}$. Now, it is easy to see that in spite of being a translationally-invariant state, if  $m$ and $n$ are of opposite parities, no quantum walk comprising of steps of the form form Eq. (\ref{eq:traditional_step}) can generate it from some home-state. \par
A more general definition of unbiased quantum steps, $\hat{\mathcal{T}}_\Delta(\delta;\boldsymbol{c})$, a quantum walk of which can realize any translationally invariant quantum state, is given in terms of its action on orthogonal pair of product states: 
\begin{widetext}
	\begin{equation}
		\begin{aligned}
			\hat{\mathcal{T}}_\Delta(\delta;\boldsymbol{c})\stat{\boldsymbol{0};g}&=\sqrt{1-\Delta}e^{i\delta}\stat{\boldsymbol{c}_\perp;g}+\sqrt{\Delta}\stat{\boldsymbol{c};g+1},\\
			\hat{\mathcal{T}}_\Delta(\delta;\boldsymbol{c})\stat{\boldsymbol{1};g}&=-\sqrt{1-\Delta}e^{-i\delta}\stat{\boldsymbol{c};g}+\sqrt{\Delta}\stat{\boldsymbol{c}_\perp;g-1}.
		\end{aligned}
		\label{eq:complex_step_action}
	\end{equation}
\end{widetext}
This quantum step is such that only a fraction $\sqrt{\Delta}$ of the amplitude participates in the walk. The remaining amplitude $\sqrt{1-\Delta}$ does not see any change in the position, but undergoes only a change of the coin-state.  
Any composite state of the form Eq. (\ref{eq:composite_state}), if it is translationally invariant (that is, satisfies Eq. (\ref{eq:TI_definition})), can be generated from some home-state using only these kind of steps \cite{kadiri2023steered}. \par
\subsection{Definition of novel quantum steps}
As discussed in the earlier section, in this paper we shall employ a new class of quantum steps, instead of $\hat{\mathcal{T}}(\alpha,\beta,\gamma)$ or $\hat{\mathcal{T}}_\Delta(\delta;\boldsymbol{c})$. We indicate these by the symbol $\hat{T}(p,q;\boldsymbol{c},\boldsymbol{s})$, where $p$ and $q$ are integers, and $\boldsymbol{s}$ and $\boldsymbol{c}$ indicate two states $\stat{\boldsymbol{s}}$ and $\stat{\boldsymbol{c}}$ on the coin-space respectively. The action of this quantum step on the orthogonal home-states is as:
\begin{equation}
	\begin{aligned}
		\hat{T}(p,q;\boldsymbol{c},\boldsymbol{s})\stat{\boldsymbol{s};g}&=\stat{\boldsymbol{c};g+p},\\
		\hat{T}(p,q;\boldsymbol{c},\boldsymbol{s})\stat{\boldsymbol{s}_\perp;g}&=\stat{\boldsymbol{c}_\perp;g+q}.
	\end{aligned}
	\label{eq:simple_step_action}
\end{equation}
This quantum step moves the walker $p$ units ahead from the current position if the coin state is $\stat{\boldsymbol{s}}$, and $q$ units ahead from the current position if the coin state is $\stat{\boldsymbol{s}_\perp}$, while simultaneously transforming the coin state to  $\stat{\boldsymbol{c}}$ and $\stat{\boldsymbol{c}_\perp}$, respectively. We shall consider Eq. (\ref{eq:simple_step_action}) itself as  the definition of the quantum step $\hat{T}(p,q;\boldsymbol{c},\boldsymbol{s})$. The action of this step on any other product state $\stat{\boldsymbol{w};g}$ can be obtained by expressing $\stat{\boldsymbol{w}}$ in the \{$\stat{\boldsymbol{s}},\stat{\boldsymbol{s}_\perp}$\} basis, and acting $\hat{T}(p,q;\boldsymbol{c},\boldsymbol{s})$ on the superposition. Likewise, the action of $\hat{T}(p,q;\boldsymbol{c},\boldsymbol{s})$ on an arbitrary composite state  $\stat{\boldsymbol{S}}$ of the form Eq. (\ref{eq:composite_state}) can be obtained by acting it linearly on each of the terms $\stat{\boldsymbol{s}_m;m}$. 
The step acts in a translationally invariant manner, since the definition of Eq. (\ref{eq:simple_step_action}) is defined identically for all $\stat{g}$. 
In the quantum step $\hat{T}(p,q;\boldsymbol{c},\boldsymbol{s})$, if one of $(p,q)$ are zero, this would correspond to a unidrectional quantum step \cite{hoyer2009faster,montero2013unidirectional,innocenti2017quantum}, where only one component participates in the walk. Here,  however, we shall assume no relation between $p$ and $q$.  We have the following relation two quantum steps:
	\begin{equation}
\hat{T}(p,q;\boldsymbol{c},\boldsymbol{s})\equiv\hat{T}(q,p;\boldsymbol{c}_\perp,\boldsymbol{s}_\perp)
\end{equation}
We call the two states $\stat{\boldsymbol{s}}$ and $\stat{\boldsymbol{c}}$  of a quantum step $\hat{T}(p,q;\boldsymbol{c},\boldsymbol{s})$ as its coin and shift states respectively. 

The quantum step $\hat{T}(p,q,\boldsymbol{c},\boldsymbol{s})$ of Eq. (\ref{eq:simple_step_action}) carries the same character as the conventional step $\hat{\mathcal{T}}(\alpha,\beta,\gamma)$ of Eq. (\ref{eq:traditional_step}), in the sense that it can be cast in the conventional form of a shift operation followed by a coin toss operator as in Eq. (\ref{eq:traditional_step}):
\begin{equation}
	\hat{T}(p,q;\boldsymbol{c},\boldsymbol{s})=\hat{S}(p,q;\boldsymbol{c})\left(\hat{\mathit{c}}(\boldsymbol{c};\boldsymbol{s})\otimes \hat{I}_{2N+1}\right)
\end{equation}
where $\hat{S}(p,q;\boldsymbol{c})$ is the shift operator that acts as in Eq. (\ref{eq:traditional_shift}):
\begin{equation}
	\begin{aligned}
		\hat{S}(p,q;\boldsymbol{c})\stat{\boldsymbol{c};g}&=\stat{\boldsymbol{c};g+p},\\
		\hat{S}(p,q;\boldsymbol{c})\stat{\boldsymbol{c}_\perp;g}&=\stat{\boldsymbol{c}_\perp;g+q}.
	\end{aligned}
	\label{eq:new_shift}
\end{equation}
The operator $\hat{\mathit{c}}(\boldsymbol{c};\boldsymbol{s})$ is the coin-toss operator which is just an SU(2) operator on the coin space, defined as:
\begin{equation}
	\hat{\mathit{c}}(\boldsymbol{c};\boldsymbol{s})=\stat{\boldsymbol{c}}\bra{\boldsymbol{s}}+\stat{\boldsymbol{c}_\perp}\bra{\boldsymbol{s}_\perp}
	\label{eq:new_coin_toss}
\end{equation}
We prove this by showing that the step $\hat{\mathcal{T}}_\Delta(\delta;\boldsymbol{c})$ can itself be realized using a pair of $\hat{T}(p,q;\boldsymbol{c},\boldsymbol{s})$ steps, as:
\begin{widetext}
	\begin{equation}
		\begin{aligned}
			\hat{\mathcal{T}}_\Delta(\delta;\boldsymbol{c})\equiv\hat{T}(1,0,\boldsymbol{c},\boldsymbol{b})\hat{T}(0,-1,\boldsymbol{0},\boldsymbol{0}), \text{ where } 
			\stat{\boldsymbol{b}}=\sqrt{\Delta}\stat{\boldsymbol{0}}-\sqrt{1-\Delta}e^{i\delta}\stat{\boldsymbol{1}}.
		\end{aligned}
	\end{equation}
\end{widetext}
This establishes that the quantum steps  introduced here are powerful enough to generate any translationally invariant composite state from some home-state. \par

\subsection{Novel quantum walks and their action on home-states}

A quantum walk $\hat{W}$ of $N$ steps is a collection of above quantum steps:
\begin{equation}
\hat{W}=\hat{T}_N\hat{T}_{N-1}\cdots\hat{T}_1,
\label{eq:walk_defintion}
\end{equation}
with $\hat{T}_i$ being the $i^{th}$ quantum step of the form Eq. (\ref{eq:simple_step_action}), $\hat{T}_i\equiv\hat{T}\left(p_i,q_i;\boldsymbol{c}_i,\boldsymbol{s}_i\right)$. We shall call quantum walks where all the steps are identical as ``homogeneous walks", and the others as ``homogeneous walks". \par
Given a quantum walk as in Eq. (\ref{eq:walk_defintion}), we associate an integer $d$ to it, computed from  $p$ and $q$ parameters of the individual steps, as:
\begin{equation}
d=\sum_{i=1}^{N}\left(p_i+q_i\right),
\label{eq:d_of_walk_definition}
\end{equation}
where $p_i$ and $q_i$ are $p$ and $q$ parameters of the $i^{th}$ quantum step $\hat{T}_i$ defined through Eq. (\ref{eq:simple_step_action}). 
Like in case of quantum steps, one could decide label quantum walk themselves as unbiased or biased forward or biased backward, depending or whether $d$ is equal to or greater than or less than $0$ respectively. 
The action of a quantum walk $\hat{W}$ on a product state $\stat{\boldsymbol{w};0}$ is to transform it into a composite state $\stat{\boldsymbol{W}}$ of the form Eq. (\ref{eq:composite_state}):
\begin{equation}
	\begin{aligned}
		\hat{W}\stat{\boldsymbol{w};0}=\stat{\boldsymbol{W}},\, \text{ and }
		\hat{W}\stat{\boldsymbol{w}_\perp;0}=\stat{(\boldsymbol{W}_\perp)_{+d}},\\
		\label{eq:walk_action}
	\end{aligned}
\end{equation}
where $d$ is given by Eq. (\ref{eq:d_of_walk_definition}), and $\stat{\boldsymbol{W}}$ is a translationally-invariant state (defined in Eq. (\ref{eq:TI_definition})). 
Recall that any translationally invariant state can be generated from some home-state by using quantum steps of the kind $\hat{\mathcal{T}}_\Delta(\delta;\boldsymbol{c})$. This is true even for the quantum steps of $\hat{T}(p,q;\boldsymbol{c},\boldsymbol{s})$. 
Actually, it is to be noted that the quantum steps $\hat{T}(p,q;\boldsymbol{c},\boldsymbol{s})$ do not give access to new states on the composite space over and above what is possible with unbiased quantum steps $\hat{\mathcal{T}}_\Delta(\delta;\boldsymbol{c})$. To emphasize, consider the composite states $\stat{\boldsymbol{W}}$ and $\stat{(\boldsymbol{W}_\perp)_{+d}}$ appearing on the RHS of Eq. (\ref{eq:walk_action}). These orthogonal composite states result from quantum walks using steps of the form $\hat{T}(p,q;\boldsymbol{c},\boldsymbol{s})$ starting from orthogonal home-states. These two composite states can also be obtained from quantum walks using steps of the form $\hat{\mathcal{T}}_\Delta(\delta;\boldsymbol{c})$, except that a single quantum walk cannot accomplish this task, unless $d=0$. \par
Moving forward, the following points can be readily  derived about quantum walks of Eq. (\ref{eq:walk_defintion}): 
\begin{enumerate}
	\item If the steps of a quantum walk are such that the coin state of one is identical to the shift state of the subsequent step, the complete walk can be interpreted as a single step whose shift state is the shift state of the first and coin state is the coin state of the last step:
	\begin{equation}
		\prod_{i=1}^{m}\hat{T}(p_i,q_i,\boldsymbol{s}_{i+1},\boldsymbol{s}_i)=\hat{T}\left(\sum_{i=1}^{N}p_i,\sum_{i=1}^{N}p_i,\boldsymbol{s}_{m+1},\boldsymbol{s}_1\right)
		\label{eq:two_step_relation}
	\end{equation}
	\item A homogeneous quantum walk comprised of a quantum step with identical coin and shift states, is another quantum step by itself:
	\begin{equation}
		{\hat{T}(p,q,\boldsymbol{u},\boldsymbol{u})}^N=\hat{T}(Np,Nq,\boldsymbol{u},\boldsymbol{u})
		\label{eq:same_same_step_power_N_relation}
	\end{equation}
	\item With quantum steps in which shift state $\stat{\boldsymbol{s}}$ and the coin state $\stat{\boldsymbol{c}}$ are orthogonal, we have the relation on the homogeneous quantum walk comprising of an even number of them:
	\begin{equation}
		{\hat{T}(p,q,\boldsymbol{u}_\perp,\boldsymbol{u})}^N=(-1)^\frac{N}{2}\hat{T}\left(\frac{N}{2}(p+q),\frac{N}{2}(p+q);\boldsymbol{u},\boldsymbol{u}\right)
		\label{eq:same_ortho_step_power_N_relation}
	\end{equation}
\end{enumerate}\par
As most generalized initial states of a quantum walk, we shall consider not product states of the form $\stat{\boldsymbol{w};0}$, but localized mixed states of the form $\hat{\rho}_{r,\boldsymbol{s}}\otimes \stat{0}\bra{0}$,
where $\hat{\rho}_{r,\boldsymbol{s}}$ is the density matrix of the coin state, Eq. (\ref{eq:mixed_st_defintion}).
We shall call the states of this form as the ``home-states''.  
The result of the walk $\hat{W}$ on the home-states is the density matrix $\hat{\rho}_{res}$:
\begin{equation}
	\hat{\rho}_{res}=\hat{W}(\hat{\rho}_{r,\boldsymbol{s}}\otimes\stat{0}\bra{0})\hat{W}^{-1}
	\label{eq:rho_res_definition}
\end{equation}
It is to be noted that $\hat{\rho}_{res}$ is a density matrix on the composite space and $\hat{\rho}_{r,\boldsymbol{s}}$ is a density matrix on the two-dimensional coin space alone.
If the home-state $\hat{\rho}_{r,\boldsymbol{s}}$ is a pure-state density matrix (that is, if $r=1$), then the resulting quantum walk density matrix has the form $\hat{\rho}_{res}=\stat{\boldsymbol{S}}\bra{\boldsymbol{S}}$, where $\stat{\boldsymbol{S}}$ is a composite pure state of the form Eq. (\ref{eq:composite_state}). 
\section{Quantum games and Parrondo's paradox in quantum walks}
\label{sec:PP_In_QW}
\subsection{A quantum game on these quantum walks}
 
Given a walk $\hat{W}$, and an arbitrary mixed coin state $\hat{\rho}_{r,\boldsymbol{s}}$ of the form Eq. (\ref{eq:mixed_st_defintion}), we use the expression $\textbf{exp}(\hat{O},\hat{W},\hat{\rho}_{r,\boldsymbol{s}})$ to indicate the expectation value $\hat{O}$ in the density matrix resulting from the action of $\hat{W}$ at the position $\stat{0}$, in a coin state whose density matrix is given by $\hat{\rho}_{r,\boldsymbol{s}}$. We call this quantity as the payoff.\par
Consider a tuple $(\hat{W},\hat{O} ,\omega)$ where  $\hat{W}$ is a quantum walk defined in Eq. (\ref{eq:walk_defintion}), $\hat{O}$ is an Hermitian operator defined in Eq. (\ref{eq:operator_O}), and $\omega$ is a real number denoting the ``target payoff". Given these, we say a home-state $\hat{\rho}_{r,\boldsymbol{s}}$ is a ``winning state'' if the payoff, that is  $\textbf{exp}(\hat{O},\hat{W},\hat{\rho}_{r,\boldsymbol{s}})$ defined in Eq. (\ref{eq:exp_mixed_state_defintion}), is greater than $\omega$. Likewise, a ``losing state'' if the payoff is less than $\omega$, and a ``tie state'' if it is equal to $\omega$. \par
Given a walk, an operator on the composite space, and a target payoff, the game here is to identify the winning home-states, if any.
With this background, we are ready to define the Parrondo paradox. 

\subsection{Parrondo's paradox in the above game}
For realizing the Parrondo's paradox in this game, we consider $m$ quantum steps $\hat{T}_1, \cdots \hat{T}_m$ of the form Eq. (\ref{eq:simple_step_action}), and using these quantum steps, we construct $m+1$ quantum walks as follows:
\begin{equation}
	\begin{aligned}
		&\hat{W}_1=[\hat{T}_1]^{n\times m},\,
		\hat{W}_2=[\hat{T}_2]^{n\times m},\,
		\cdots,
		\hat{W}_m=[\hat{T}_m]^{n\times m}, \\
		&\text{ and } \hat{W}_{m+1}=[\hat{T}_m\cdots\hat{T}_2\hat{T}_1]^n.
	\end{aligned}
	\label{eq:PP_Walk_Scheme}
\end{equation}
The first $m$ of these walks are all homogeneous walks, while the last one is an inhomogenous walk comprising of all the steps in a sequence. Now, a home-state $\hat{\rho}_{r,\boldsymbol{s}}$  may be a winning state for some of the $m+1$ walks, a losing for some, and possibly a tie state for some other of these $m+1$ walks. 
For observing Parrondo's paradox in the above setting, we seek  a home-state $\hat{\rho}_{r,\boldsymbol{s}}$ such that it is a losing state for all the homogeneous walks $\hat{W}_i$,  for $i=1\cdots m$, but happens to be a winning state for the inhomogenous walk $\hat{W}_{m+1}$. That is 
\begin{equation}
	\begin{aligned}
	\textbf{exp}(\hat{O},\hat{W}_i,\hat{\rho}_{r,\boldsymbol{s}})&<\omega, \text{ for } i=1,\cdots,m, \text{ and }\\
	\textbf{exp}(\hat{O},\hat{W}_{m+1},\hat{\rho}_{r,\boldsymbol{s}})&>\omega.
	\end{aligned}
	\label{eq:Parrondo_State_definition}
\end{equation}
We call the home-states $\hat{\rho}_{r,\boldsymbol{s}}$ that satisfy the Eq. (\ref{eq:Parrondo_State_definition}) as ``Parrondo states''. 
It is to be noted that the walks of Eq. (\ref{eq:PP_Walk_Scheme}) contain a positive integer parameter $n$. We call this $n$ as the number of cycles. It is possible that a pure or mixed home-state is a Parrondo state for a certain $n=n_0$ but not for any $n<n_0$ and $n>n_0$. Now, we say a Parrondo state $\hat{\rho}_{r,\boldsymbol{s}}$ to be a ``persistent Parrondo state'' if it satisfies Eq. (\ref{eq:Parrondo_State_definition}) for all $n$ greater than a certain $n_0$. 

\section{Parrondo's paradox in arbitrary quantum walks}
\label{sec:PP_in_ArbWalks}
By definition, the payoff $\textbf{exp}(\hat{O},\hat{W},\hat{\rho}_{r,\boldsymbol{s}})$ can be computed as
\begin{equation}
	\textbf{exp}(\hat{O},\hat{W},\hat{\rho}_{r,\boldsymbol{s}})\equiv \textbf{tr}(\hat{W}^{-1}\hat{O}\hat{W}(\hat{\rho}_{r,\boldsymbol{s}}\otimes\stat{0}\bra{0})),
	\label{eq:exp_mixed_state_defintion}
\end{equation}
where $\textbf{tr}$ stands for trace operator over the composite system. 
If the home-state density matrix $\hat{\rho}_{r,\boldsymbol{s}}$ is a pure state $\stat{\boldsymbol{s}}\bra{\boldsymbol{s}}$, then we indicate the expectation value on the LHS of Eq. (\ref{eq:exp_mixed_state_defintion}) as $\textbf{exp}(\hat{O},\hat{W},\boldsymbol{s})$, and it simplifies to:
\begin{equation}
	\textbf{exp}(\hat{O},\hat{W},\boldsymbol{s})\equiv \ip{\boldsymbol{s};0|\hat{W}^{-1}\hat{O}\hat{W}}{\boldsymbol{s};0}
	\label{eq:exp_purestate_defintion}
\end{equation}
Now, it is straight-forward to see that a convex combination of Parrondo states will also be a Parrondo state.  We will therefore search for Parrondo states only among the pure states.
\par
We shall first try to compute the expression for the payoff $\textbf{exp}(\hat{O},\hat{W},\boldsymbol{s})$ for a given home-state $\stat{\boldsymbol{s}}$. 
\subsection{Home-states leading to the largest expectation values}
Given a walk $\hat{W}$ and an operator $\hat{O}$, we shall refer to the minimum and maximum payoffs possible by $o^{max}$ and $o_{min}$, and the coin-states that lead to these payoffs by the symbols $\stat{\boldsymbol{o}^{max}}$ and $\stat{\boldsymbol{o}_{min}}$ respectively:
\begin{equation}
	\begin{aligned}
	\stat{\boldsymbol{o}_{max}}&=\textbf{max}_{\textbf{exp}(\hat{O},\hat{W},\boldsymbol{s})}\left\{{\stat{\boldsymbol{s}}}\right\},\\
	o^{max}&=\textbf{exp}(\hat{O},\hat{W},\boldsymbol{o}^{max}),\\
	\stat{\boldsymbol{o}_{min}}&=\textbf{min}_{\textbf{exp}(\hat{O},\hat{W},\boldsymbol{s})}\left\{\stat{\boldsymbol{s}}\right\},\\
	o_{min}&=\textbf{exp}(\hat{O},\hat{W},\boldsymbol{o}_{min}).
	\end{aligned}
\end{equation}
In other words, this coin-states $\stat{\boldsymbol{o}^{max}}$ and $\stat{\boldsymbol{o}_{min}}$ are such that $\textbf{exp}(\hat{O},\hat{W},\boldsymbol{o}^{max})\ge \textbf{exp}(\hat{O},\hat{W},\hat{\rho}_{r,\boldsymbol{s}})$ and $\textbf{exp}(\hat{O},\hat{W},\boldsymbol{o}_{min})\le \textbf{exp}(\hat{O},\hat{W},\hat{\rho}_{r,\boldsymbol{s}})$ for all density matrices $\hat{\rho}_{r,\boldsymbol{s}}$ of the coin-space.  It is evident that the maximum and minimum payoffs can be expected only with pure states, not mixed states.
\par
First we will try to identify these $\stat{\boldsymbol{o}^{max}}$ and $\stat{\boldsymbol{o}_{min}}$ coin states, which, for a given walk $\hat{W}$ and operator $\hat{O}$, are unique up to a global phase.
These states are actually the eigenstates of a Hermitian operator $\hat{o}$, defined as:
\begin{equation}
	\hat{o}= \textbf{tr}_{walk}\left((\hat{I}_2\otimes \stat{0}\bra{0})\hat{W}^{-1}\hat{O}\hat{W}\right),
	\label{eq:small_o_operator}
\end{equation}
where $ \textbf{tr}_{walk}$ refers to the tracing over the walk degree-of-freedom. The resulting operator $\hat{o}$ is a $2\times2$ Hermitian operator on the coin space. Note that $\hat{o}$ depends on the quantum walk $\hat{W}$, in addition to the operator $\hat{O}$.
In an arbitrary basis $\{\stat{\boldsymbol{w};\boldsymbol{w}_\perp}\}$ of the coin space, the matrix representation of $\hat{o}$ is given by:
\begin{equation}
	\hat{o}=\left[\begin{array}{cc}
		\ip{\boldsymbol{W}}{\hat{O}|\boldsymbol{W}}&  \ip{\boldsymbol{W}}{\hat{O}|(\boldsymbol{W}_\perp)_{+d}}\\
		\ip{(\boldsymbol{W}_\perp)_{+d}}{\hat{O}|\boldsymbol{W}}& \ip{(\boldsymbol{W}_\perp)_{+d}}{\hat{O}|(\boldsymbol{W}_\perp)_{+d}}
	\end{array}\right],
\end{equation}
where $\stat{\boldsymbol{W}}$ and $\stat{(\boldsymbol{W}_\perp)_{+d}}$ are the composite states obtained by action of the walk $\hat{W}$ on home-states $\stat{\boldsymbol{w};0}$ and  $\stat{\boldsymbol{w}_\perp;0}$ respectively. (see Eq. (\ref{eq:walk_action})). 
The desired $o^{max}$ and $o_{min}$ are the larger and smaller eigenvalues of the matrix $\hat{o}$ respectively, and the states $\stat{\boldsymbol{o}^{max}}$ and $\stat{\boldsymbol{o}_{min}}$ are the corresponding eigenvectors. We have $\ip{\boldsymbol{o}^{max}}{\boldsymbol{o}_{min}}=0$, since $\hat{o}$ is a Hermitian operator. \par
The payoff possible with any other home-state $\stat{\boldsymbol{s}}$ can be can be computed directly from these quantities as:

\begin{equation}
	\begin{aligned}
		\textbf{exp}(\hat{O},\hat{W},\boldsymbol{u})=&\frac{1}{2}\left((o^{max}-o_{min})\vect{S}(\boldsymbol{o}^{max})\cdot\vect{S}(\boldsymbol{u})\right)\\
		+&\frac{1}{2}(o^{max}+o_{min}).
		\label{eq:expectation_analytic}
	\end{aligned}
\end{equation}
Here the $\vect{S}(\boldsymbol{o}^{max})$ and $\vect{S}(\boldsymbol{u})$ are the qubit vectors (defined in Eq. (\ref{eq:stokes_vector_pure_states})) of the coin-states $\stat{\boldsymbol{o}^{max}}$ and $\stat{\boldsymbol{u}}$ respectively. Likewise, the payoff in case of mixed-states can also be computed as:
\begin{equation}
	\begin{aligned}
		\textbf{exp}(\hat{O},\hat{W},\hat{\rho}_{r,\boldsymbol{s}})=&\frac{1}{2}\left((2r-1)(o^{max}-o_{min})\vect{S}(\boldsymbol{o}^{max})\cdot\vect{S}(\boldsymbol{s})\right)\\
		+&\frac{1}{2}(o^{max}+o_{min}).
	\end{aligned}
\end{equation}
The maximally mixed state corresponds to $r=\frac{1}{2}$, and for this case the expectation value is $\frac{1}{2}(o^{max}+o_{min})$.
\subsection{Winning the quantum game}
Recall that we say a home-state $\stat{\boldsymbol{u};0}$ is a winning state if its payoff  $\textbf{exp}(\hat{O},\hat{W},\boldsymbol{u})$, is greater than the target payoff $\omega$. Now, for a given pair of $\hat{W}$ and $\hat{O}$, it is possible that the minimum and maximum possible payoffs are equal, in which case it follows from Eq. (\ref{eq:expectation_analytic}) that every home-state yields the same payoff. This would be mean the whole Bloch sphere behaves identically, that is, the whole sphere is a winning or losing or tie region, depending on the target payoff. Otherwise, we define a quantity $\Omega$ as follows:
\begin{equation}
	\Omega=\frac{2\omega-(o^{max}+o_{min})}{o^{max}-o_{min}}
	\label{eq:Omega_definition}
\end{equation}
Therefore, from Eq. (\ref{eq:expectation_analytic}), it follows that a pure home-state $\stat{\boldsymbol{u};0}$ is a winning state if and only if the dot-product $\vect{S}(\boldsymbol{o}^{max})\cdot\vect{S}(\boldsymbol{u})>\Omega$, and it is a losing state if and only if $\vect{S}(\boldsymbol{o}^{max})\cdot\vect{S}(\boldsymbol{u})<\Omega$. Home-states with $\vect{S}(\boldsymbol{o}^{max})\cdot\vect{S}(\boldsymbol{u})=\Omega$, are the ``tie-states". 
Since the winning criteria of a home-state $\stat{\boldsymbol{u}}$ depends only on the dot product of its qubit vector and the qubit vector of $\stat{\boldsymbol{o}^{max}}$,
We shall now try to locate these Parrondo states in a unit-sphere called the Bloch sphere, where every state $\hat{\rho}_{r,\boldsymbol{s}}$ is represented by its qubit vector $\vect{S}(\hat{\rho}_{r,\boldsymbol{s}})$.  Pure states are located on surface of this sphere and the mixed states are placed within, with the maximally mixed state placed at centre of the sphere. 
It follows that if the home-state $\stat{\boldsymbol{s}}$ is in the $\stat{\boldsymbol{o}^{max}}$ and $\stat{\boldsymbol{o}_{min}}$ basis as
\begin{equation}
	\stat{\boldsymbol{s}}=\cos\frac{\nu}{2}\stat{\boldsymbol{o}^{max}}+e^{i\delta}\sin\frac{\nu}{2}\stat{\boldsymbol{o}_{min}},
\end{equation}
All home-states with the same latitude $\nu$ have the same expectation values. 
It follows therefore that the tuple $(\hat{W},\hat{O} ,\omega)$ splits the Bloch sphere into two contiguous regions, such that the states in one region are all winning states and those in the other are all losing states. The states along the circle separating these spherical caps are the tie states. The normals to this circle are the qubit vectors $\vect{S}(\boldsymbol{o}^{max})$ and $\vect{S}(\boldsymbol{o}_{min})$. If $\Omega=0$, that is if the target payoff $\omega$ is taken to be equal to $\frac{1}{2}(o^{max}+o_{min})$, then the two spherical caps become two hemispheres, and the circle separating them will be a great-circle. 
\subsection{Identifying the Parrondo states on the Bloch sphere}
That was about single quantum walk. Now, for identifying Parrondo states, we shall have to deal with $m+1$ walks of Eq. (\ref{eq:PP_Walk_Scheme}). A given state $\stat{\boldsymbol{u}}$ is a winning state for the walk $\hat{W}_i$ if the dot product $\vect{S}(\boldsymbol{o}^{max}_i).\vect{S}(\boldsymbol{u})$ is greater than $\Omega_i$. Here $\Omega_i$ is the $\Omega$ of Eq. (\ref{eq:Omega_definition}), computed for the walk $\hat{W}_i$. The surface of the Bloch sphere thus splits into utmost $2^{m+1}$ disjoint slices, depending upon whether each state $\stat{\boldsymbol{u}}$ on it is a winning state for $\hat{W}_i$ or not. Whether all $2^{m+1}$ distinct regions exist depend upon the choice of the target payoff $\omega$, and also on the $m+1$ stokes vectors $\vect{S}(\boldsymbol{o}^{max}_i)$. 
One slice among these, for instance, might be comprising of home-states  which are winning for all $m+1$ walks simultaneously, and likewise another region might be corresponding to home-states which are losing states simultaneously for all of them. One among these regions corresponds to all home-states that are losing states for the $m$ walks $\hat{W}_i, i=1,\cdots,m$ but are winning states for the $\hat{W}_{m+1}$ walk. These are the Parrondo states that we are after. \par
It must be understood that biased quantum walks or biased quantum steps are not essential for the above scheme of identifying the Parrondo's state to work. In other words, the steps $\hat{T}_i$ of Eq. (\ref{eq:PP_Walk_Scheme}) could be the steps $\hat{\mathcal{T}}_\Delta(\delta;\boldsymbol{c})$ of Eq. (\ref{eq:complex_step_action}) or $\hat{\mathcal{T}}(\alpha,\beta,\gamma)$ of Eq. (\ref{eq:traditional_step}) as well.\par
\subsection{Persistence of Parrondo's paradox}
Consider the quantum walk $\hat{W}_i$, one of the $m+1$ quantum walks of Eq. (\ref{eq:PP_Walk_Scheme}). Recall that these quantum walks are defined for a particular cycle-length $n$. We shall refer to the walk $\hat{W}_i$ defined for a particular $n$ by the symbol $\hat{W}_i(n)$. For a given operator $\hat{O}$ and the quantum walk $\hat{W}_i(n)$, we shall indicate the coin-space operator $\hat{o}$ of Eq. (\ref{eq:small_o_operator}) by the symbol $\hat{o}_i(n)$. The Parrondo regions corresponding to the cycle-lengths $n$ and $n'$ will be identical if the matrices $\hat{o}_i(n)$  and $\hat{o}_i(n')$ commute for all $i=1,\cdots,m+1$.In the general case, however, the matrices do not commute. In that case, the intersection of the Parrondo regions correspond to $n$ and $n'$ would  lead to home-state that display Parrondo's effect at both the cycle-lengths. 
The intersection of Parrondo regions corresponding to all $n>n_0$ for some $n_0$ comprise of the persistent Parrondo states.  \par We shall give an illustration of this in Section \ref{sub_sec:PP_in_Arb_Walk}.
\section{Parrondo's paradox with biased quantum steps}
\label{sec:Accentuate_PP}
The above construction of identifying Parrondo's paradox works for any kind of quantum steps, biased or unbiased. The composite states that result from these quantum walks are entangled, spanning multiple position states. The manifestation of Parrondo paradox in these quantum walks can be attributed to the interference affects of probability amplitudes. This raises an interesting question: are entanglement and delocalized probability amplitudes indispensable for the emergence of this paradox? In this section, we answer this question in the negative. We show that one could design quantum steps such that a desired home-state happens to be the Parrondo state for the mean-position operator. These quantum steps are such that the chosen home-state coherently hops on the position state, without being in the superposition of position states at any time. The collection of quantum steps can be thought of as  displaying ``classical Parrondo's paradox''. The construction here however, works only with biased quantum steps, and for only certain operators. 
\subsection{Zero-position operator}
In this illustration, we consider the Hermitian operator $\hat{O}$ to be $\hat{0}=\hat{I}_2\otimes \stat{0}\bra{0}$. 
The expectation value of $\hat{0}$ in a composite state $\stat{\boldsymbol{S}}$ of the form Eq. (\ref{eq:composite_state}), is $s_0^2$. 
We can take the target payoff $\omega$ to be any number strictly between 0 and 1: $0<\omega<1$. For a quantum walk $\hat{W}$, a home state $\stat{\boldsymbol{u};0}$ is a winning state if in the resulting composite state $\hat{W}\stat{\boldsymbol{u};0}$, the probability at position zero is greater than $\omega$. 
Here we demonstrate that two quantum steps, where a walk of 
Here we consider two quantum steps of the form:
\begin{equation}
	\hat{T}_A=\hat{T}(m,m,\boldsymbol{c}_1,\boldsymbol{s}_1),\, \text{ and }\hat{T}_B=\hat{T}(-m,-m,\boldsymbol{c}_2,\boldsymbol{s}_2)
\end{equation}
for some $m\neq 0$.
From these we construct four quantum walks as 
\begin{equation}
	\begin{aligned}
		\hat{W}_A=[\hat{T}_A]^{2n},\,\hat{W}_B=[\hat{T}_B]^{2n},\\
		\hat{W}_{AB}=[\hat{T}_B\hat{T}_A]^n,\text{ and }\hat{W}_{BA}=[\hat{T}_A\hat{T}_B]^n.
	\end{aligned}
\end{equation}
Using these two steps, we shall demonstrate the Parrondo paradox,  simultaneously for two triplets of walks: $(\hat{W}_A,\hat{W}_B,\hat{W}_{AB})$ and $(\hat{W}_A,\hat{W}_B,\hat{W}_{BA})$.\par
Consider a walker starting from a home-state $\stat{\boldsymbol{u};0}$. The walks $\hat{W}_A$ and $\hat{W}_B$ place the walker at the locations $\stat{2nm}$ and $\stat{-2nm}$  respectively, in some coin-states.
The walks  $\hat{W}_{AB}$ and $\hat{W}_{BA}$, on the other hand, are just SU(2) transformations on the coin-space alone, and do not change the position of the walker. Therefore the walker remains at home, $\stat{0}$, albeit in a different coin state, when acted on either of them. Therefore we have
\begin{equation}
	\begin{aligned}
	\textbf{exp}(\hat{I_2}\otimes\hat{0},\hat{W}_A,\boldsymbol{s})&=\textbf{exp}(\hat{I_2}\otimes\hat{0},\hat{W}_B,\boldsymbol{s})=0,\\
	\textbf{exp}(\hat{I_2}\otimes\hat{0},\hat{W}_{AB},\boldsymbol{s})&=\textbf{exp}(\hat{I_2}\otimes\hat{0},\hat{W}_{BA},\boldsymbol{s})=1.
	\end{aligned}
\end{equation}
Therefore for the operator $\hat{0}$, together with any target payoff $0<\omega<1$, every state on the Bloch sphere is a losing state for the walks $\hat{W}_A$ and $\hat{W}_B$,  but is a winning state for both the walks, $\hat{W}_{AB}$ and $\hat{W}_{BA}$. 
The complete Bloch sphere is a Parrondo region. Furthermore, this result holds true for any $n$, leading to the whole Bloch sphere being a persistent Parrondo region. 
\subsection{Mean position operator}
\label{sec:mean_pos_operator}
In this section, we shall consider the operator $\hat{O}$ of Eq. (\ref{eq:exp_mixed_state_defintion}) to be $\hat{I}_2\otimes\hat{\mu}$, where $\hat{\mu}$ is the position operator as
\begin{equation}
	\hat{\mu}=\sum_{m=-N}^{N}m\stat{m}\bra{m}
	\label{eq:mean_position_operator}
\end{equation}
The expectation value of $\hat{\mu}$ in a composite state $\boldsymbol{S}$ is given by 
\begin{equation}
\ip{\boldsymbol{S}}{\hat{I}_2\otimes\hat{\mu}|\boldsymbol{S}}=\sum_{m=b}^{e}ms_m^2.
\end{equation}
We have the following points regarding this position operator, which we shall need further down the text. 
\begin{enumerate}
	\item \label{enum:ortho_relation} For a given quantum walk $\hat{W}$ and two orthogonal pure states, we have the following relationship between the mean position they generate:
$\textbf{exp}(\hat{I_2}\otimes\hat{\mu},\hat{W},\boldsymbol{s}_\perp)=d-\textbf{exp}(\hat{I_2}\otimes\hat{\mu},\hat{W},\boldsymbol{s})$, with $d$ of the walk $\hat{W}$ defined is given by Eq. (\ref{eq:d_of_walk_definition}). 
	\item \label{enum:both_neg} If $p$ and $q$ parameters of a quantum step $\hat{T}$ are both negative, then in a homogeneous walk of $N$ such steps will never lead to a positive mean value for any home-state. That is, $\textbf{exp}\left(\hat{I}_2\otimes\hat{\mu},\hat{T}^N(p,q;\boldsymbol{c},\boldsymbol{s}),\hat{\rho}_{r,\boldsymbol{s}}\right)<0$ if $p<0$ and $q<0$. 
\end{enumerate}
We shall take the target payoff $\omega$ to be equal to $0$. This means that a home state $\hat{\rho}_{r,\boldsymbol{s}}$ is a winning state if the walk $\hat{W}$ transforms $\hat{\rho}_{r,\boldsymbol{s}}\otimes\stat{0}\bra{0}$ to a composite density matrix $\hat{\rho}_{res}$ having a positive mean position. \par

In this subsection, we will explicitly create a set of quantum steps such that a given home state $\stat{\boldsymbol{w};0}$ is a Parrondo state.  
Consider a collection of even number $m$ of quantum steps, defined as 
\begin{equation}
	\begin{aligned}
	\hat{T}_1&=\hat{T}(p_1,q_1,\boldsymbol{c}_1,\boldsymbol{w}),\\
	\hat{T}_2&=\hat{T}(p_2,q_2,\boldsymbol{c}_2,\boldsymbol{c}_1),\\
	&\cdots,\\
	\hat{T}_{m-1}&=\hat{T}(p_{m-1},q_{m-1},\boldsymbol{w}_\perp,\boldsymbol{c}_{m-2}),\\	
	\hat{T}_m&=\hat{T}(p_m,q_m,\boldsymbol{w},\boldsymbol{w}_\perp).\\	
	\end{aligned}
	\label{eq:pp_construction}
\end{equation}
Here $\stat{\boldsymbol{c}_i} \text{ for } i=1,\cdots,m-2$ and $\stat{\boldsymbol{w}}$ are arbitrary coin-states. These quantum steps are ``daisy-chained'', in the sense that the coin state of the $i^{th}$ step is the shift state of $(i+1)^{th}$ step. The step sizes $p_i$ and $q_i$ must satisfy the following conditions: 
\begin{equation}
	\begin{aligned}
	p_i&<0\,\forall i=1,\cdots, m-1, \,  \text{ and } p_m>-\sum_{i=1}^{m-1}p_i,\\q_i&<0\,\forall i=1,\cdots, m, \\
	q_m&<\sum_{i=1}^{m-1}p_i, \text{ and }\, p_m<-q_m.
	\end{aligned}
	\label{eq:PP_Design}
\end{equation}
\begin{figure}[!ht]
	\centering
	\includegraphics[height=\textheight,width=\linewidth,keepaspectratio]{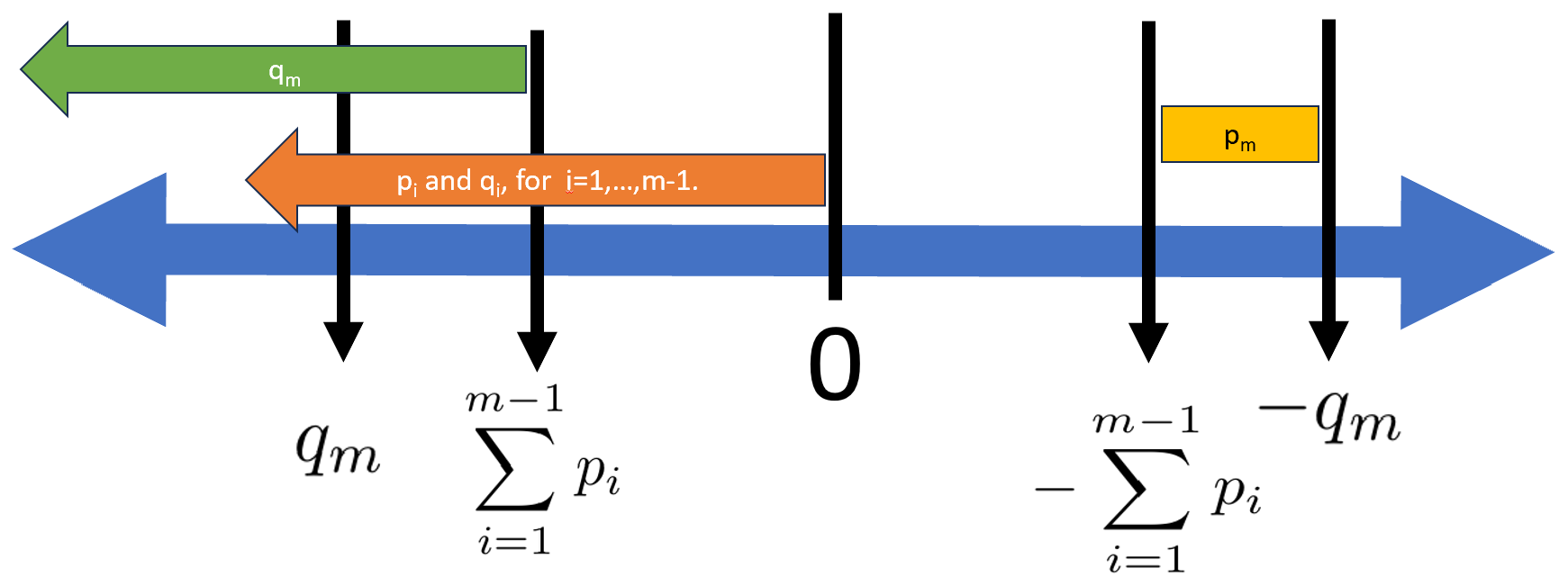}
	\caption{ Constraints on the step sizes of the quantum steps. The step sizes $p_i \text{ and }q_i, \text{ for } i=1,\cdots,m-1$ are all negative. The step size $\hat{q}_m$ should be more negative than $\sum_{i=1}^{m-1}p_i$. The step-size $p_m$ is positive, and must lie between two the non-negative integers, $-\sum_{i=1}^{m-1}p_i$ and $-q_m$.}
	\label{fig:stepsizes}
\end{figure}

We shall now examine if the home-state $\stat{\boldsymbol{w}}$ is a Parrondo state for the collection of walks of the form Eq. (\ref{eq:PP_Walk_Scheme})  constructed out of these steps. The state $\stat{\boldsymbol{w};0}$ will be a losing state for the $m-1$ walks $\hat{W}_i$ for $i=1,\cdots,m-1$ following the point (\ref{enum:both_neg}) above. Consider now the $m^{th}$ quantum walk, $\hat{W}_m$. Since $m$ is taken to be an even number, we have from Eq. (\ref{eq:same_ortho_step_power_N_relation}) the following action of the walk $\hat{W}_m$:
\begin{equation}
\hat{W}_m=\hat{T}\left(\frac{nm}{2}(p_m+q_m),\frac{n m}{2}(p_m+q_m),\boldsymbol{w},\boldsymbol{w}\right)
\end{equation}
Since $p_m+q_m<0$ (see Eq. (\ref{eq:PP_Design})), the walk $\hat{W}_m$ is also a losing walk for all $n$. \par
Consider now the walk $\hat{W}_{m+1}=[\prod_{i=m}^{1}\hat{T}_i]^n$. From Eq.(\ref{eq:two_step_relation}) and Eq. (\ref{eq:same_same_step_power_N_relation}), this walk is:
\begin{equation}
	\hat{W}_{m+1}=\hat{T}\left(n\sum_{i=1}^{m}p_i,n\sum_{i=1}^{m}q_i;\boldsymbol{w},\boldsymbol{w}\right)
	\label{eq:W_m+1}
\end{equation}
Since $\sum_{i=1}^{m}p_i>0$, the home-state $\stat{\boldsymbol{w};0}$ is a winning state for $\hat{W}_{m+1}$. This home-state therefore satisfies Eq. (\ref{eq:Parrondo_State_definition}), and therefore is a Parrondo state. \\
Having demonstrated that $\stat{\boldsymbol{w}}$ is a Parrondo state for this set of quantum walks, we shall now see if these quantum walks support any other Parrondo state. 
Indeed the $m$ quantum walks $\{\hat{W}_1,\cdots,\hat{W}_m\}$ are, by design, such that every state $\stat{\boldsymbol{s}}$ on the Bloch sphere is a losing state for all of them. Therefore a home-state $\stat{\boldsymbol{s}}$ is a Parrondo state if it happens to be a winning state for the walk $\hat{W}_{m+1}$. For identifying the winning states of the walk $\hat{W}_{m+1}$, consider its action on the home-state $\stat{\boldsymbol{s}}$. We have from the definition of this walk, Eq. (\ref{eq:W_m+1}), the following:
\begin{equation}
	\hat{W}_{m+1}\stat{\boldsymbol{s};0}=\ip{\boldsymbol{w}}{\boldsymbol{s}}\stat{\boldsymbol{w};n\sum_{i=1}^{m}p_i}+\ip{\boldsymbol{w}_\perp}{\boldsymbol{s}}\stat{\boldsymbol{w}_\perp;n\sum_{i=1}^{m}q_i}
\end{equation}
The expectation value of the operator $\hat{\mu}$ in the composite state of the RHS is given by:
\begin{equation}
	\textbf{exp}(\hat{\mu},\hat{W}_{m+1},\boldsymbol{s})=n\sum_{i=1}^{m}\left(q_i+\abs{\ip{\boldsymbol{w}}{\boldsymbol{s}}}^2(p_i-q_i)\right)
\end{equation}
This expectation value is positive if 
\begin{equation}
	\abs{\ip{\boldsymbol{w}}{\boldsymbol{s}}}^2>\frac{\sum_{i=1}^{m}q_i}{\sum_{i=1}^{m}(q_i-p_i)}.
\end{equation}
Therefore all states $\stat{\boldsymbol{s}}$ on the Bloch sphere, of the form $\stat{\boldsymbol{s}}=\cos\frac{\nu}{2}\stat{\boldsymbol{w}}+e^{i\delta}\sin\frac{\nu}{2}\stat{\boldsymbol{w}_\perp}$, with $\nu<\nu_{max}$ are Parrondo states, where $\nu_{max}=\cos^{-1}\left(\frac{2\sum_{i=1}^{m}q_i}{\sum_{i=1}^{m}(q_i-p_i)}\right)$. Furthermore, all Parrondo states are persistent Parrondo states.

\section{Numerical illustrations}
\label{sec:Numerical_Illustration}
In this section we shall provide some illustrations of the schemes discussed above. 
We shall define a few coin states:
\begin{equation}
	\begin{aligned}
		\stat{\boldsymbol{h}}&=\frac{1}{\sqrt{2}}\left(\stat{\boldsymbol{0}}+\stat{\boldsymbol{1}}\right),\, 
		\stat{\boldsymbol{v}}=\frac{1}{\sqrt{2}}\left(\stat{\boldsymbol{1}}-\stat{\boldsymbol{0}}\right),\\
		\stat{\boldsymbol{d}}&=\frac{1}{\sqrt{2}}\left(\stat{\boldsymbol{0}}+i\stat{\boldsymbol{1}}\right),\, 
		\stat{\boldsymbol{a}}=\frac{1}{\sqrt{2}}\left(i\stat{\boldsymbol{0}}+\stat{\boldsymbol{1}}\right),\\
		\stat{\boldsymbol{f}}&=\cos\left(\frac{\pi}{8}\right)\stat{\boldsymbol{0}}+\sin\left(\frac{\pi}{8}\right)\stat{\boldsymbol{1}}.
	\end{aligned}
	\label{eq:some_coin_states}
\end{equation}
We first provide two illustrations for the construction proposed in section \ref{sec:mean_pos_operator}, and then give an illustration for the Parrondo paradox in arbitrary quantum walks, discussed in section \ref{sec:PP_in_ArbWalks}. 
\subsection{Quantum walks that accentuate the Parrondo's Paradox}
Here we shall demonstrate the construction of Section (\ref{sec:mean_pos_operator}) for constructing the quantum steps such that a desired home-state is a Parrondo state for the mean-position operator $\hat{\mu}$.
\subsubsection{A two-step example}
In this example, we intend to have the home-state $\stat{\boldsymbol{h}}$ defined in Eq. (\ref{eq:some_coin_states}) as the Parrondo state. We take $m$ of Eq. (\ref{eq:PP_Walk_Scheme}) to be $2$, and take the two steps as:
\begin{equation}
	\begin{aligned}
		\hat{T}_1=\hat{T}(-1,-1,\boldsymbol{v},\boldsymbol{h}),\, \text{ and }
		\hat{T}_2=\hat{T}(3,-4,\boldsymbol{h},\boldsymbol{v}),
	\end{aligned}
	\label{eq:two_step_example}
\end{equation}
The chosen steps are in the form listed in Eq. (\ref{eq:pp_construction}), and satisfy the conditions of Eq. (\ref{eq:PP_Design}). 
Using these two quantum steps, we define three quantum walks:
\begin{equation}
		\hat{W}_1\equiv[\hat{T}_1]^{2n},\,
		\hat{W}_2\equiv[\hat{T}_2]^{2n}\, \text{ and }
		\hat{W}_3\equiv[\hat{T}_2\hat{T}_1]^n.
\end{equation}
The progression of the home state $\stat{\boldsymbol{h};0}$ through these three walks is depicted (up to a global phase) in Eq. (\ref{eq:First_example_flow}), for  $n=1$.
\begin{widetext}
\begin{equation}
	\begin{aligned}
		&\hat{W}_1:\;\stat{\boldsymbol{h};0}\rightarrow\hat{T}_1\rightarrow\stat{\boldsymbol{v};-1}\rightarrow\hat{T}_1\rightarrow\stat{\boldsymbol{h};-2}.\\
		&\hat{W}_2:\;\stat{\boldsymbol{h};0}\rightarrow\hat{T}_2\rightarrow\stat{\boldsymbol{v};-4}\rightarrow\hat{T}_2\rightarrow\stat{\boldsymbol{h};-1}.\\
		&\hat{W}_3:\;\stat{\boldsymbol{h};0}\rightarrow\hat{T}_1\rightarrow\stat{\boldsymbol{v};-1}\rightarrow\hat{T}_2\rightarrow\stat{\boldsymbol{h};2}. 
	\end{aligned}
	\label{eq:First_example_flow}
\end{equation}
\end{widetext}
Since $\stat{\boldsymbol{h};0}$ is a losing state for both $\hat{W}_1$ and $\hat{W}_2$ while being a winning state for $\hat{W}_3$, it is a Parrondo state, as desired. Furthermore, it is also evident that $\stat{\boldsymbol{h};0}$ is a persistent Parrondo state. \par 
To see whether any other home-states are Parrondo states, we find the action of these quantum walks on an arbitrary home-state $\stat{\boldsymbol{s}}$. We note the following:
\begin{equation}
	\begin{aligned}
		\hat{W}_1\stat{\boldsymbol{s};0}&=\stat{\boldsymbol{s};-2n},\,\hat{W}_2\stat{\boldsymbol{s};0}=\stat{\boldsymbol{s};-n},\,\\
		\hat{W}_3\stat{\boldsymbol{s};0}&=\ip{\boldsymbol{h}}{\boldsymbol{s}}\stat{\boldsymbol{h};2n}+\ip{\boldsymbol{v}}{\boldsymbol{s}}\stat{\boldsymbol{v};-5n}.
	\end{aligned}
\end{equation}
The expectation value of $\hat{\mu}$ in the three composite states $\hat{W}_i\stat{\boldsymbol{s};0}$ for the three walks are therefore:
\begin{equation}
	\begin{aligned}
		\textbf{exp}(\hat{\mu},\hat{W}_1,\boldsymbol{s})&=-2n,\,
		\textbf{exp}(\hat{\mu},\hat{W}_2,\boldsymbol{s})=-n,\\
		\textbf{exp}(\hat{\mu},\hat{W}_3,\boldsymbol{s})&=(7\abs{\ip{\boldsymbol{h}}{\boldsymbol{s}}}^2-5)n.
	\end{aligned}
\end{equation}
Evidently, home-state $\stat{\boldsymbol{s}}$ is a losing state for $\hat{W}_1$ and $\hat{W}_2$, while for $\hat{W}_3$ it is a winning state if $\abs{\ip{\boldsymbol{h}}{\boldsymbol{s}}}^2> \frac{5}{7}$. The mean position varies proportional to $n$ in all the three cases, and hence a Parrondo state will be a persistent Parrondo state. 
\subsubsection{A four-step example}
As a second example, we consider $m=4$, and the desired Parrondo state as $\stat{\boldsymbol{0}}$.  We take the four steps as following:
\begin{equation}
	\begin{aligned}
	\hat{T}_1&=\hat{T}(-1,-1,\boldsymbol{h},\boldsymbol{0}),\,
	\hat{T}_2=\hat{T}(-1,-1,\boldsymbol{d},\boldsymbol{h}),\\
	\hat{T}_3&=\hat{T}(-1,-1,\boldsymbol{1},\boldsymbol{d}), \text{ and }
	\hat{T}_4=\hat{T}(4,-5,\boldsymbol{0},\boldsymbol{1}).\\
	\end{aligned}
	\label{eq:four_step_example}
\end{equation}
As in the previous example, with these four quantum steps we define five walks: 
\begin{equation}
	\hat{W}_i=[{\hat{T}_i}]^{4n},\, i=1,2,3,4 \text{ and } \hat{W}_5=[\hat{T}_4\hat{T}_3\hat{T}_2\hat{T}_1]^n.
	\label{eq:second_example_walks}
\end{equation}
The progression of home-state $\stat{\boldsymbol{0};0}$ through each of the steps of these walks is shown (up to a global phase of the coin-state) in Eq. (\ref{eq:second_example_flow}) for $n=1$.
\begin{widetext}
	\begin{equation}
		\begin{aligned}
			&\hat{W}_1:\;\stat{\boldsymbol{0};0}\rightarrow\hat{T}_1\rightarrow\stat{\boldsymbol{h};-1}\rightarrow\hat{T}_1\rightarrow\stat{\boldsymbol{1};-2}\rightarrow\hat{T}_1\rightarrow\stat{\boldsymbol{v};-3}\rightarrow\hat{T}_1\rightarrow\stat{\boldsymbol{0};-4}.\\
			&\hat{W}_2:\;\stat{\boldsymbol{0};0}\rightarrow\hat{T}_2\rightarrow\stat{\boldsymbol{v};-1}\rightarrow\hat{T}_2\rightarrow\stat{\boldsymbol{a};-2}\rightarrow\hat{T}_2\rightarrow\stat{\boldsymbol{0};-3}\rightarrow\hat{T}_2\rightarrow\stat{\boldsymbol{v};-4}.\\
			&\hat{W}_3:\;\stat{\boldsymbol{0};0}\rightarrow\hat{T}_3\rightarrow\stat{\boldsymbol{a};-1}\rightarrow\hat{T}_3\rightarrow\stat{\boldsymbol{0};-2}\rightarrow\hat{T}_3\rightarrow\stat{\boldsymbol{a};-3}\rightarrow\hat{T}_3\rightarrow\stat{\boldsymbol{0};-4}.\\
			&\hat{W}_4:\;\stat{\boldsymbol{0};0}\rightarrow\hat{T}_4\rightarrow\stat{\boldsymbol{1};-5}\rightarrow\hat{T}_4\rightarrow\stat{\boldsymbol{0};-1}\rightarrow\hat{T}_4\rightarrow\stat{\boldsymbol{1};-6}\rightarrow\hat{T}_4\rightarrow\stat{\boldsymbol{0};-2}.\\
			&\hat{W}_5:\;\stat{\boldsymbol{0};0}\rightarrow\hat{T}_1\rightarrow\stat{\boldsymbol{h};-1}\rightarrow\hat{T}_2\rightarrow\stat{\boldsymbol{d};-2}\rightarrow\hat{T}_3\rightarrow\stat{\boldsymbol{1};-3}\rightarrow\hat{T}_4\rightarrow\stat{\boldsymbol{0};1}.
		\end{aligned}
		\label{eq:second_example_flow}
	\end{equation}
\end{widetext}
Evidently the home-state $\stat{\boldsymbol{0}}$ is a losing state for the walks $\hat{W}_i, \, i=1,\cdots,4$, but is a winning state for $\hat{W}_5$, and therefore is a Parrondo state. To examine if these walks support any other Parrondo state, we just need to examine the action of the fifth walk $\hat{W}_5$ (of Eq. (\ref{eq:second_example_walks})) on an arbitrary home-state $\stat{\boldsymbol{s};0}$. The walk $\hat{W}_5$ is actually equal to $\hat{T}(n,-8n;\boldsymbol{0},\boldsymbol{0})$, and therefore we have:
\begin{equation}
	\hat{W}_5\stat{\boldsymbol{s};0}=\ip{\boldsymbol{0}}{\boldsymbol{s}}\stat{\boldsymbol{0};n}+\ip{\boldsymbol{1}}{\boldsymbol{s}}\stat{\boldsymbol{1};-8n},
\end{equation}
and therefore we have
\begin{equation}
	\textbf{exp}(\hat{\mu},\hat{W}_5,\boldsymbol{s})=n(9\abs{\ip{\boldsymbol{0}}{\boldsymbol{s}}}^2-8).
\end{equation}
A home-state $\stat{\boldsymbol{s}}$ is a winning state for $\hat{W}_5$, and hence a Parrondo state for the quantum steps of Eq. (\ref{eq:four_step_example}), if $\abs{\ip{\boldsymbol{0}}{\boldsymbol{s}}}>\frac{2\sqrt{2}}{3}$.
\subsection{Parrondo's paradox in arbitrary quantum walks}
Here we give two illustrations for elucidating the procedure for identifying the Parrondo states, given in Section \ref{sec:PP_in_ArbWalks}. Here we shall construct walks of the form \ref{eq:PP_Walk_Scheme}.
\label{sub_sec:PP_in_Arb_Walk}
\subsubsection{A two-step illustration}
\label{sec:two_step_PP}
Here we shall take $m$ of Eq. (\ref{eq:PP_Walk_Scheme}) to be $m=2$, and consider the following two unbiased quantum steps: 
\begin{equation}
	\begin{aligned}
		\hat{T}_1=\hat{T}(1,-1;\boldsymbol{f},\boldsymbol{0}),\, \text{ and }
		\hat{T}_2=\hat{T}(1,-1;\boldsymbol{d},\boldsymbol{f}),
	\end{aligned}
	\label{eq:two_steps_PP}
\end{equation}
where $\stat{\boldsymbol{f}}$ and $\stat{\boldsymbol{h}}$ are coin-states defined in Eq. (\ref{eq:some_coin_states}). 
These steps are such that $\hat{T}_2\hat{T}_1$ is itself another quantum step $\hat{T}_2\hat{T}_1=\hat{T}(2,-2;\boldsymbol{d},\boldsymbol{0})$.\\

With these two steps, we construct three walks as in Eq. (\ref{eq:PP_Walk_Scheme}), with $n=3$:
\begin{equation}
	\begin{aligned}
	\hat{W}_1&=\hat{T}_1^6,\,
	\hat{W}_2=\hat{T}_2^6,\, \\ \text{ and }
	\hat{W}_3&=\hat{T}_2\hat{T}_1\hat{T}_2\hat{T}_1\hat{T}_2\hat{T}_1.
	\label{eq:three_walks}
	\end{aligned}
\end{equation}
For this illustration, we take the Hermitian operator to be the mean-position operator $\hat{\mu}$, defined in Eq. (\ref{eq:mean_position_operator}) as earlier. 
\begin{figure}[!ht]
	\centering
	\includegraphics[height=\textheight,width=\linewidth,keepaspectratio]{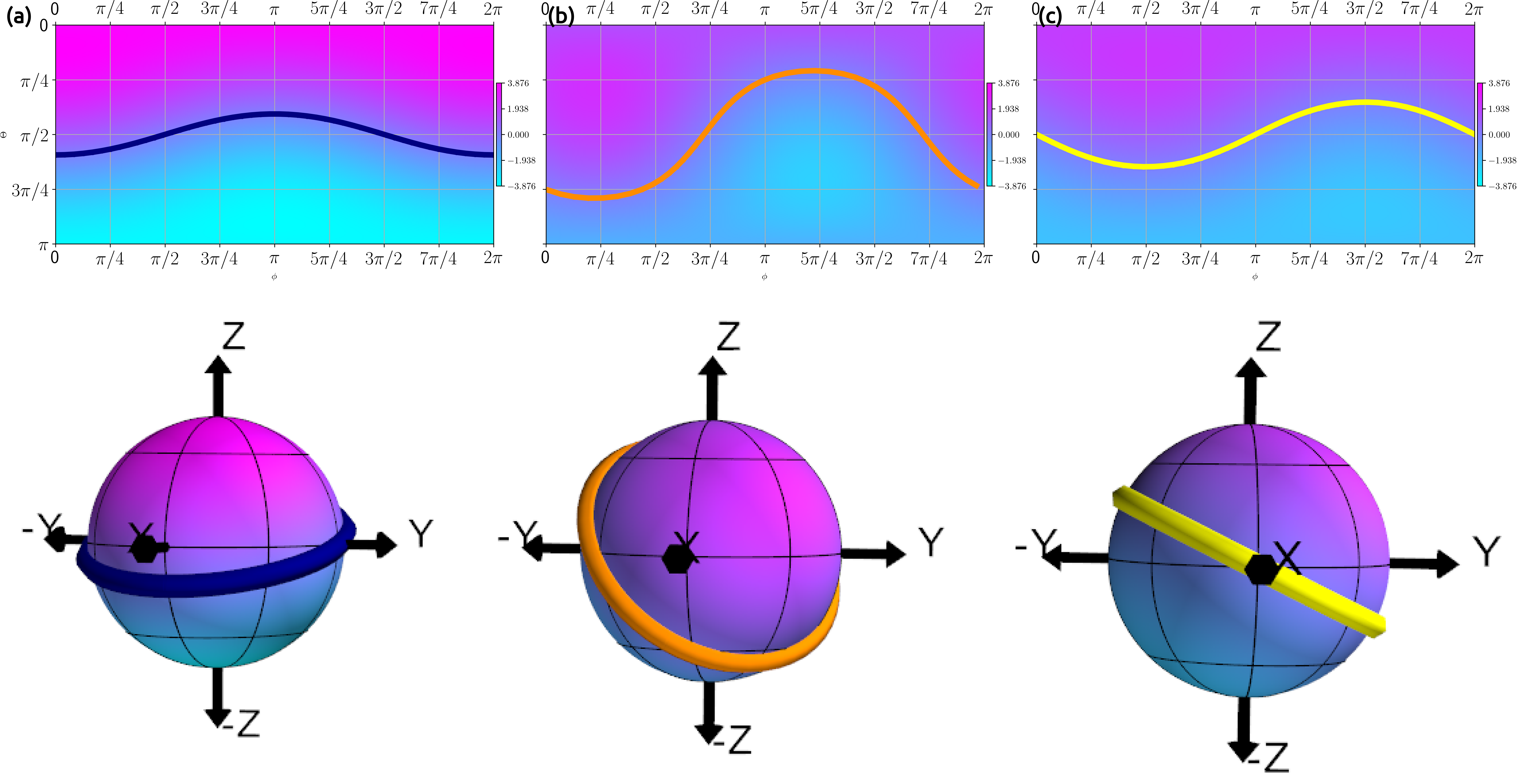}
	\caption{Mean position of the composite state obtained of the Bloch sphere into eight regions, corresponding to three walks defined in Eq. (\ref{eq:three_walks}). The operator being position operator $\hat{\mu}$. The navy-blue, orange and yellow arrow indicate the Bloch vector direction of the coin states $\stat{\boldsymbol{o^{max}}}$ of the three walks respectively.}
	\label{fig:PP_Bloch_Sp_mu_operator}
\end{figure}
\begin{figure}[!ht]
	\centering
	\includegraphics[height=\textheight,width=\linewidth,keepaspectratio]{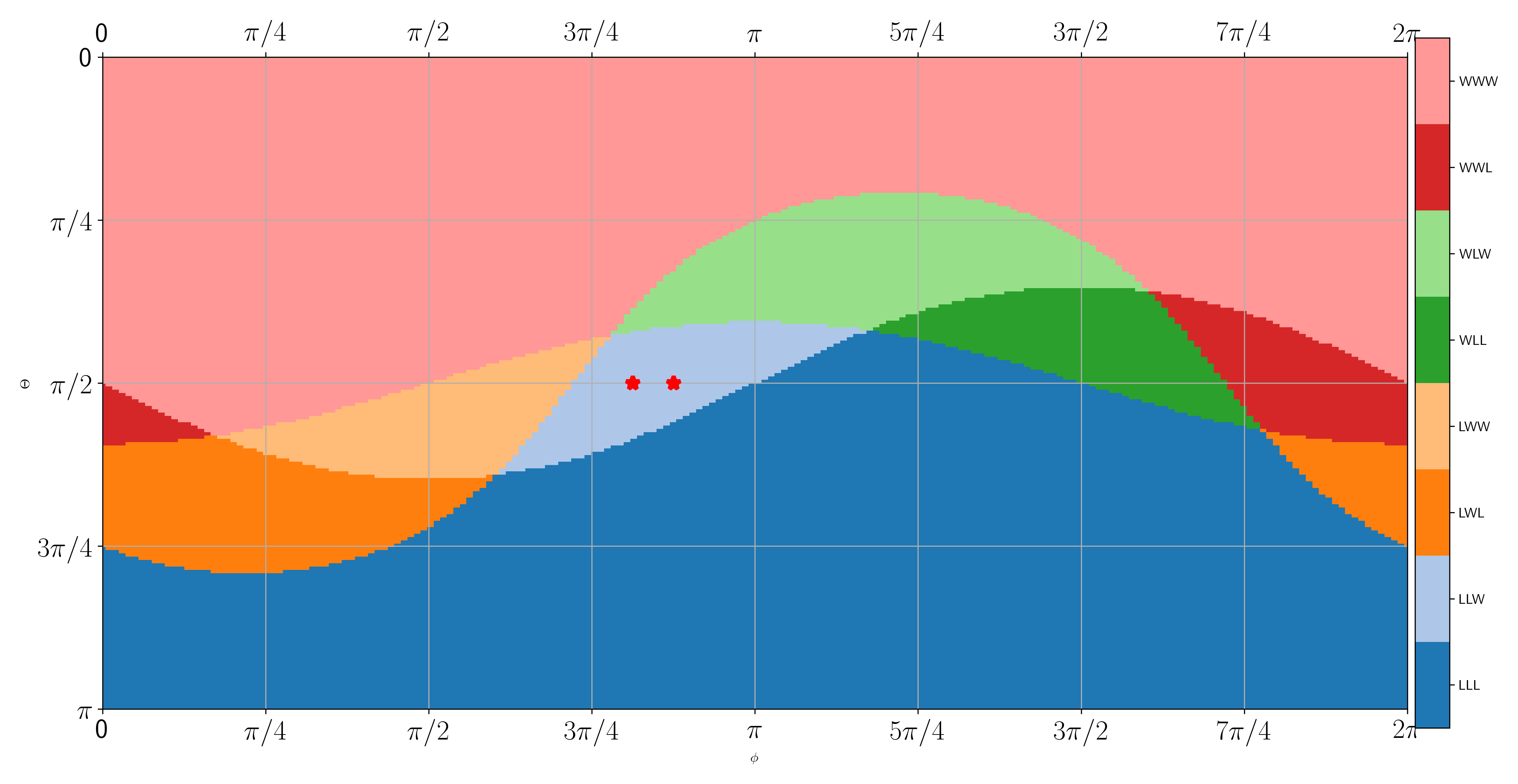}
	\caption{Identification of eight regions on the $(\theta,\phi)$ color-coded plot. These regions correspond to home-states that lead to different winning and losing combinations, corresponding to the three walks of Eq. (\ref{eq:three_walks}), and for the mean-position operator $\hat{\mu}$.}
	\label{fig:PP_MeanPoS_WinningLosing2d}
\end{figure}
In the basis $\stat{\boldsymbol{0}}=(1,0)^T$ and  $\stat{\boldsymbol{1}}=(0,1)^T$, the $2\times 2$ matrices for the operators $\hat{o}_i$ are given:
\begin{equation}
	\begin{aligned}
		\hat{o}_1&=\left[\begin{array}{cc}
			3.71& 1.123\\
			1.123&-3.71 
		\end{array}\right],\\
		\hat{o}_2&=\left[\begin{array}{cc}
			1.503& 1.503-1.25i\\
			1.503+1.25i&-1.503 
		\end{array}\right],\\
		\hat{o}_3&=\left[\begin{array}{cc}
			2& -i\\
			i&-2 
		\end{array}\right].
	\end{aligned}
	\label{eq:mu_eg_matrices}
\end{equation}
The eigen-vectors of these matrices can be found to be:
\begin{table}
	\begin{tabular}{|c|c|c|}
		\hline
		&$\stat{\boldsymbol{o}^{max}}$  &$\stat{\boldsymbol{o}_{min}}$  \\
		\hline
		$\hat{o}_1$&$0.989\stat{\boldsymbol{0}}+0.146\stat{\boldsymbol{1}}$  &$0.146\stat{\boldsymbol{0}}-0.989\stat{\boldsymbol{1}}$  \\
		\hline
		$\hat{o}_2$&$0.897\stat{\boldsymbol{0}}+(0.34+0.282i)\stat{\boldsymbol{1}}$  &$(0.34-0.282i)\stat{\boldsymbol{0}}-0.897\stat{\boldsymbol{1}}$   \\
		\hline
		$\hat{o}_3$&$0.973\stat{\boldsymbol{0}}+0.23i\stat{\boldsymbol{1}}$  &$0.23\stat{\boldsymbol{0}}-0.973i\stat{\boldsymbol{1}}$   \\
		\hline
	\end{tabular}
	\caption{Eigenvectors $\stat{\boldsymbol{o}^{max}}$ and $\stat{\boldsymbol{o}_{min}}$ of the operator $\hat{o}_i$ of Eq. (\ref{eq:mu_eg_matrices}).}
	\label{tab:eigen_vector_mean_pos}
\end{table}
The corresponding eigen-values are given in Table \ref{tab:eigen_values_mean_pos} below.

\begin{table}
	\begin{tabular}{|c|c|c|c|}
		\hline
		&$\hat{o}_1$  &$\hat{o}_2$  &$\hat{o}_3$  \\
		\hline
		$o^{max}$&3.876  &2.465  &2.236  \\
		\hline
		$o_{min}$&-3.876  &-2.465  &-2.236  \\
		\hline
	\end{tabular}
	\caption{Eigenvalues of the operator $\hat{o}_i$ of Eq. (\ref{eq:mu_eg_matrices}).}
	\label{tab:eigen_values_mean_pos}
\end{table}

One can identify two states in the Parrondo region as:
\begin{equation}
\stat{\boldsymbol{\psi}_1}=\frac{1}{\sqrt{2}}\left(\stat{\boldsymbol{0}}+e^{i\frac{13\pi}{16}}\stat{\boldsymbol{1}}\right), \stat{\boldsymbol{\psi}_2}=\frac{1}{\sqrt{2}}\left(\stat{\boldsymbol{0}}+e^{i\frac{7\pi}{8}}\stat{\boldsymbol{1}}\right)
\label{eq:pp_example_states}
\end{equation}
These two states are indicated by two star shaped markers in Fig. (\ref{fig:PP_MeanPoS_WinningLosing2d}). 
One could also realize a mixed Parrondo state by mixing these two states as:
\begin{equation}
	\hat{\rho}_{12}=\frac{1}{2}\stat{\boldsymbol{\psi}_1}\bra{\boldsymbol{\psi}_1}+\frac{1}{2}\stat{\boldsymbol{\psi}_2}\bra{\boldsymbol{\psi}_2}
	\label{eq:mixed_state_example}
\end{equation}

The mean-positions of the corresponding composite states for the three walks $\stat{\boldsymbol{\psi}_1}$, $\stat{\boldsymbol{\psi}_2}$ and $\hat{\rho}$ are displayed in Table. \ref{tab:mean_pos_PP_table}. Evidently, all three home-states are losing states for $\hat{W}_1$ and $\hat{W}_2$ but are winning states for $\hat{W}_3$, establishing that these are Parrondo states. \par
We now examine the Parrondo's paradox for the same three walks of Eq. (\ref{eq:three_walks}), but for a different observable $\hat{\Delta}$, also defined on the position-space alone:

\begin{table}
\begin{tabular}{|c|c|c|c|}
	\hline
	&  $\hat{W}_1$&$\hat{W}_2$  &  $\hat{W}_3$\\
	\hline
	$\stat{\boldsymbol{\psi}_1}$&-0.934  &-0.555 &0.556  \\	\hline
	$\stat{\boldsymbol{\psi}_2}$&-1.038  &-0.91  &0.383  \\
	\hline
	$\hat{\rho}_{12}$&-0.986  &-0.732  & 0.469 \\
	\hline
\end{tabular}
\caption{Three examples of Parrondo states. Mean position of the composite state resulting from the three walks of Eq. (\ref{eq:three_walks}), starting from the pure home-states $\stat{\boldsymbol{\psi}_1}$ and $\stat{\boldsymbol{\psi}_2}$ of Eq. (\ref{eq:pp_example_states}), and the mixed home-state $\hat{\rho}_{12}$ of Eq. (\ref{eq:mixed_state_example})}
\label{tab:mean_pos_PP_table}
\end{table}
\begin{equation}
\hat{\Delta}=\sum_{m=1}^{N}\stat{m}\bra{m}-\sum_{m=-1}^{-N}\stat{m}\bra{m}
\label{eq:Delta_definition}
\end{equation}
The expectation value of $\hat{\Delta}$ in a pure state $\stat{\boldsymbol{S}}$ gives the difference between the probabilities of occupying positive and negative positions:
\begin{equation}
		\ip{\boldsymbol{S}}{\hat{I}_2\otimes\hat{\Delta}|\boldsymbol{S}}=\sum_{m>0}^{}s_m^2-\sum_{m<0}^{}s_m^2,
\end{equation}
The coin-space operator $\hat{o}$ of Eq. (\ref{eq:small_o_operator}), corresponding to this observable, can be computed by taking $\hat{O}$ to be $\hat{I}_2\otimes\hat{\Delta}$.
The three operators corresponds to the three walks are:
\begin{equation}
	\begin{aligned}
		\hat{o}_1&=\left[\begin{array}{cc}
			0735& 0.304\\
			0.304&-0.735 
		\end{array}\right],\\
		\hat{o}_2&=\left[\begin{array}{cc}
			0.353&0.353-0.375i\\
			0.353+0.375i&-0.353 
		\end{array}\right],\\
		\hat{o}_3&=\frac{1}{2}\left[\begin{array}{cc}
			1& -i\\
			i&-1 
		\end{array}\right].
	\end{aligned}
	\label{eq:delta_eg_matrices}
\end{equation}

The  two orthogonal eigen-states $\stat{\boldsymbol{o}^{max}}$ and $\stat{\boldsymbol{o}_{min}}$ of $\hat{o}$, for each of the operators $\hat{o}_i$ of Eq. (\ref{eq:delta_eg_matrices}) are tabulated in Table \ref{tab:eigenstates_prob_diff}, and the corresponding eigenvalues $o^{max}$ and $o_{min}$ are tabulated in Table \ref{tab:eigenvalues_prob_diff}.

\begin{table}
	\begin{tabular}{|c|c|c|}
		\hline
		&$\stat{\boldsymbol{o}^{max}}$  &$\stat{\boldsymbol{o}_{min}}$  \\
		\hline
		$\hat{o}_1$&$0.981\stat{\boldsymbol{0}}+0.195\stat{\boldsymbol{1}}$  &$0.195\stat{\boldsymbol{0}}-0.981\stat{\boldsymbol{1}}$  \\
		\hline
		$\hat{o}_2$&$0.885\stat{\boldsymbol{0}}+(0.32+0.34i)\stat{\boldsymbol{1}}$  &$(0.32-0.282i)\stat{\boldsymbol{0}}-0.885\stat{\boldsymbol{1}}$   \\
		\hline
		$\hat{o}_3$&$0.924\stat{\boldsymbol{0}}+0.383i\stat{\boldsymbol{1}}$  &$0.383\stat{\boldsymbol{0}}-0.924i\stat{\boldsymbol{1}}$   \\
		\hline
	\end{tabular}
	\caption{Eigenvectors $\stat{\boldsymbol{o}^{max}}$ and $\stat{\boldsymbol{o}_{min}}$  of the operators $\hat{o}_i$ of Eq. (\ref{eq:delta_eg_matrices}).}
	\label{tab:eigenstates_prob_diff}
\end{table}

\begin{table}
	\begin{tabular}{|c|c|c|c|}
		\hline
		&$\hat{o}_1$  &$\hat{o}_2$  &$\hat{o}_3$  \\
		\hline
		$o^{max}$&0.796  &0.625 &0.7071  \\
		\hline
		$o_{min}$&-0.796 &-0.625&-0.7071\\
		\hline
	\end{tabular}
	\caption{Eigenvalues of the operators $\hat{o}_i$ of Eq. (\ref{eq:delta_eg_matrices}).}
	\label{tab:eigenvalues_prob_diff}
\end{table}

\begin{figure}[!ht]
	\centering
	\includegraphics[height=\textheight,width=\linewidth,keepaspectratio]{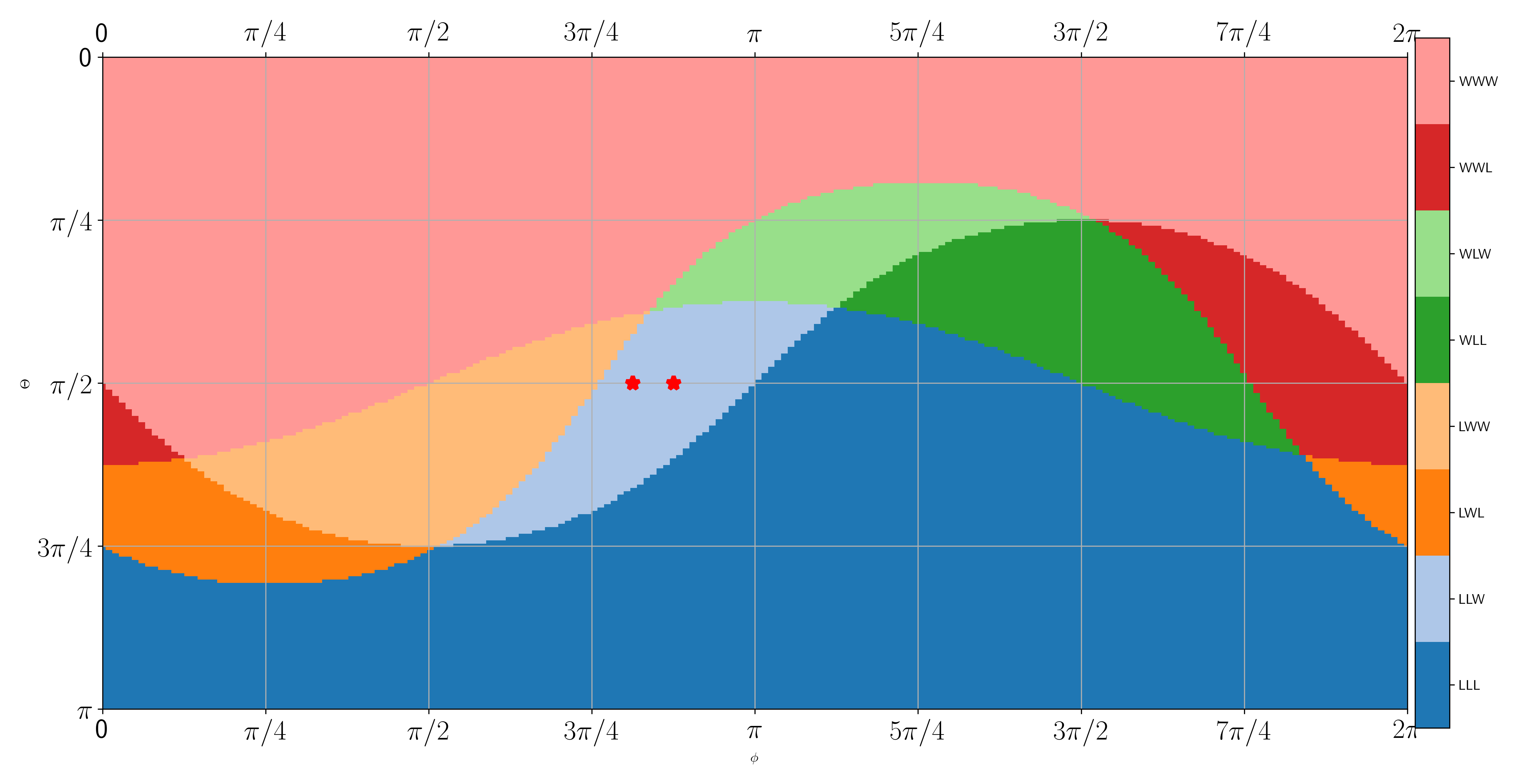}
	\caption{$(\theta,\phi)$ plot for identify the eight regions on the Bloch sphere that lead to different winning and losing combinations corresponding to the three walks of Eq. (\ref{eq:three_walks}) for the probability difference operator $\hat{\Delta}$ of Eq. (\ref{eq:Delta_definition}). }
	\label{fig:PP_ProbDIff_WinningLosing2d}
\end{figure}

The eight regions on the Bloch sphere can be identified as shown in Fig. (\ref{fig:PP_ProbDIff_WinningLosing2d}). This plot looks very similar but is not identical to the corresponding figure for the mean position operator $\hat{\mu}$, Fig. (\ref{fig:PP_MeanPoS_WinningLosing2d}). 
The two states $\stat{\boldsymbol{\psi}_1}$ and $\stat{\boldsymbol{\psi}_2}$ of Eq. (\ref{eq:pp_example_states}) which were chosen arbitrarily from the Parrondo region of the mean position operator $\hat{\mu}$, fall in the Parrondo region of the operator $\hat{\Delta}$ too. The payoffs $\textbf{exp}(\hat{I}_2\otimes\hat{\Delta},\hat{W},\boldsymbol{\psi})$ for $\stat{\boldsymbol{\psi}}$ being $\stat{\boldsymbol{\psi}_1}$ and $\stat{\boldsymbol{\psi}_2}$ of Eq. (\ref{eq:pp_example_states}), and the payoff $\textbf{exp}(\hat{I}_2\otimes\hat{\Delta},\hat{W},\hat{\rho}_{12})$ for the density matrix $\hat{\rho}_{12}$ of Eq. (\ref{eq:mixed_state_example}), for the three walks $\hat{W}=\hat{W}_1,\hat{W}_2$ and $\hat{W}_3$ are tabulated in Table. \ref{tab:prob_diff_PP_table}. 

\begin{table}
	\begin{tabular}{|c|c|c|c|}
		\hline
		&  $\hat{W}_1$&$\hat{W}_2$  &  $\hat{W}_3$\\
		\hline
		$\stat{\boldsymbol{\psi}_1}$&-0.253  &-0.086  & 0.278 \\	\hline
		$\stat{\boldsymbol{\psi}_2}$&-0.281  &-0.183 &0.191\\
		\hline
		$\hat{\rho}_{12}$& -0.267 & -0.134 & 0.235 \\
		\hline
	\end{tabular}
	\caption{Three examples of Parrondo states. Difference between the probability of occupation of positive and negative positions of the composite states resulting from the  three walks of Eq. (\ref{eq:three_walks}), starting from three home-states: the pure states $\stat{\boldsymbol{\psi}_1}$ and $\stat{\boldsymbol{\psi}_2}$ and the mixed state $\hat{\rho}_{12}$ of Eq. (\ref{eq:mixed_state_example})}
	\label{tab:prob_diff_PP_table}
\end{table}

\subsubsection*{Persistence of this Parrondo Paradox}
We will now examine three Parrondo states $\stat{\psi_1},\stat{\psi_1}$ of Eq. (\ref{eq:pp_example_states}) and $\hat{\rho}_{12}$ of Eq. (\ref{eq:mixed_state_example}) for their persistence. 
Consider the two quantum steps of Eq. (\ref{eq:two_step_example}). From these, we construct three walks $\hat{W}_1(n)=\hat{T}_1^{2n}$, $\hat{W}_2(n)=\hat{T}_2^{2n}$ and $\hat{W}_3(n)=[\hat{T}_2\hat{T}_1]^{n}$. The six panels $a-c$ of Fig. (5) represent the payoffs $\textbf{exp}(\hat{\mu},\hat{W},\psi_1)$, $\textbf{exp}(\hat{\mu},\hat{W},\psi_2)$, and $\textbf{exp}(\hat{\mu},\hat{W},\hat{\rho}_{12})$ respectively, and the panels $d-f$ of the same figure depict the payoffs $\textbf{exp}(\hat{\Delta},\hat{W},\psi_1)$,
$\textbf{exp}(\hat{\Delta},\hat{W},\psi_2)$, and
$\textbf{exp}(\hat{\Delta},\hat{W},\hat{\rho}_{12})$ respectively. The orange, green and yellow plots in each of the six panels correspond to the walks $\hat{W}$ being $\hat{W}_1(n)$, $\hat{W}_2(n)$ and $\hat{W}_3(n)$ respectively. We have plotted these payoffs as a function of $n=1$ to $19$. It is visible that in all the six cases the yellow plot stays above zero the orange and blue plots remain below. This demonstrates that the three coin-states $\stat{\psi_1},\stat{\psi_2}$ and $\hat{\rho}_{12}$ are persistent Parrondo states. 
This is shown .
\par in Fig. (\ref{fig:PP_2Step_Histogram}) we show the composite states associated with the home-state $\stat{\psi_1}$  three walks (a) $\hat{T}_1^{2n}$, (b) $\hat{T}_2^{2n}$ and (c) $[\hat{T}_2\hat{T}_1]^n$ for $n=1,2,3$ and $4$.  We depict a composite state $\stat{\boldsymbol{S}}=\sum_{m=b}^{m=e}s_m\stat{\boldsymbol{s}_m;m}$ as a horizontal histogram. Towards this, we mark the walk positions $\stat{m}$ along the y-axis, and place at every $\stat{m}$, a horizontal bar of length proportional to the quantity $s_m^2$. The histogram bars corresponding to negative $m$ are colored red, while those of positive $m$  are colored green, and that corresponding to $s_0^2$ is shown in black. 

\begin{figure}[!ht]
	\centering
	\includegraphics[height=\textheight,width=\linewidth,keepaspectratio]{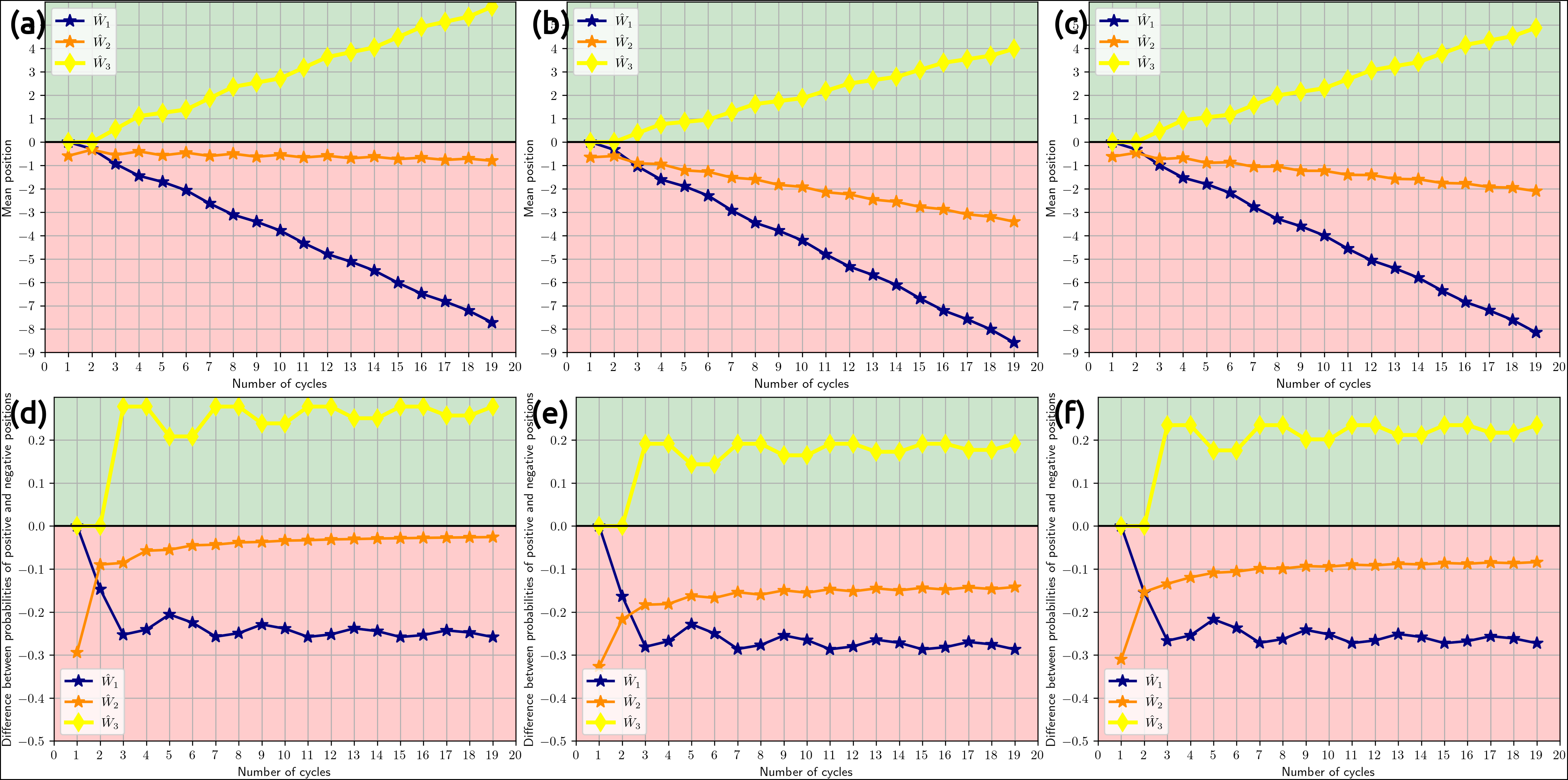}
	\caption{For $n=1$ to $n=20$. Demonstration of persistence of Parrondo's paradox, for three home-states, the two pure states $\stat{\boldsymbol{\psi}_1}$ and $\stat{\boldsymbol{\psi}_2}$ of Eq. (\ref{eq:pp_example_states}) and the mixed state $\rho_{12}$ of Eq. (\ref{eq:mixed_state_example}), for the three walks of Eq. (\ref{eq:three_walks}). Top and bottom rows correspond to the operators  $\mu$ and $\hat{\Delta}$ respectively. The navy-blue, orange and yellow colored curves correspond to the walks $\hat{W}_1$, $\hat{W}_2$ and $\hat{W}_3$ respectively. Columnwise, the columns from left to right corresponds $\stat{\boldsymbol{\psi}_1}$, $\stat{\boldsymbol{\psi}_2}$ and $\hat{\rho}_{12}$ respectively.}
	\label{fig:Persistent_PP}
\end{figure}
\begin{figure*}[!ht]
	\centering
	\includegraphics[height=\textheight,width=\linewidth,keepaspectratio]{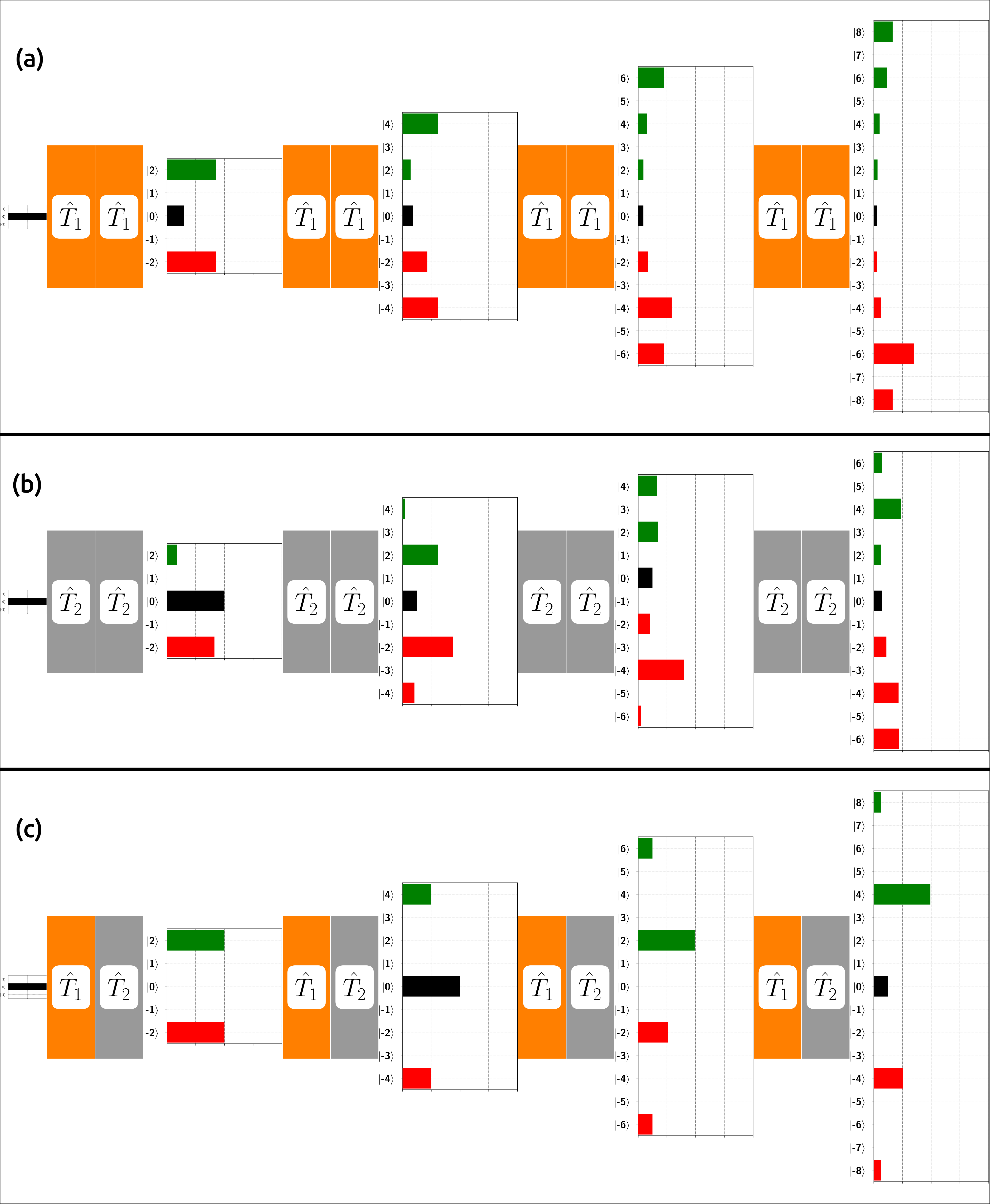}
	\caption{Histogram of probabilities at different walk positions, for the home-state $\stat{\boldsymbol{\psi}_1}$ of Eq. (\ref{eq:pp_example_states}) for the three walks (a) $\hat{T}_1^{2n}$, (b) $\hat{T}_2^{2n}$ and (c) $[\hat{T}_2\hat{T}_1]^n$ for $n=1,2,3$ and $4$, with $\hat{T}_1$ and $\hat{T}_2$ defined in Eq. (\ref{eq:two_steps_PP}).}
	\label{fig:PP_2Step_Histogram}
\end{figure*}

\subsubsection{A three step illustration}
Here we shall provide yet another illustration for identifying Parrondo regions, this time using three quantum steps. Consider the following three steps:
\begin{equation}
	\begin{aligned}
		\hat{T}_1&=\hat{T}(1,-2,\boldsymbol{h},\boldsymbol{l}),\\
		\hat{T}_2&=\hat{T}(1,-1,\boldsymbol{f},\boldsymbol{d}), \text{ and }\\
		\hat{T}_3&=\hat{T}(2,-1,\boldsymbol{f},\boldsymbol{d}).\\
	\end{aligned}
	\label{eq:Three_steps_PP}
\end{equation}

\begin{figure}[!ht]
	\centering
	\includegraphics[height=\textheight,width=\linewidth,keepaspectratio]{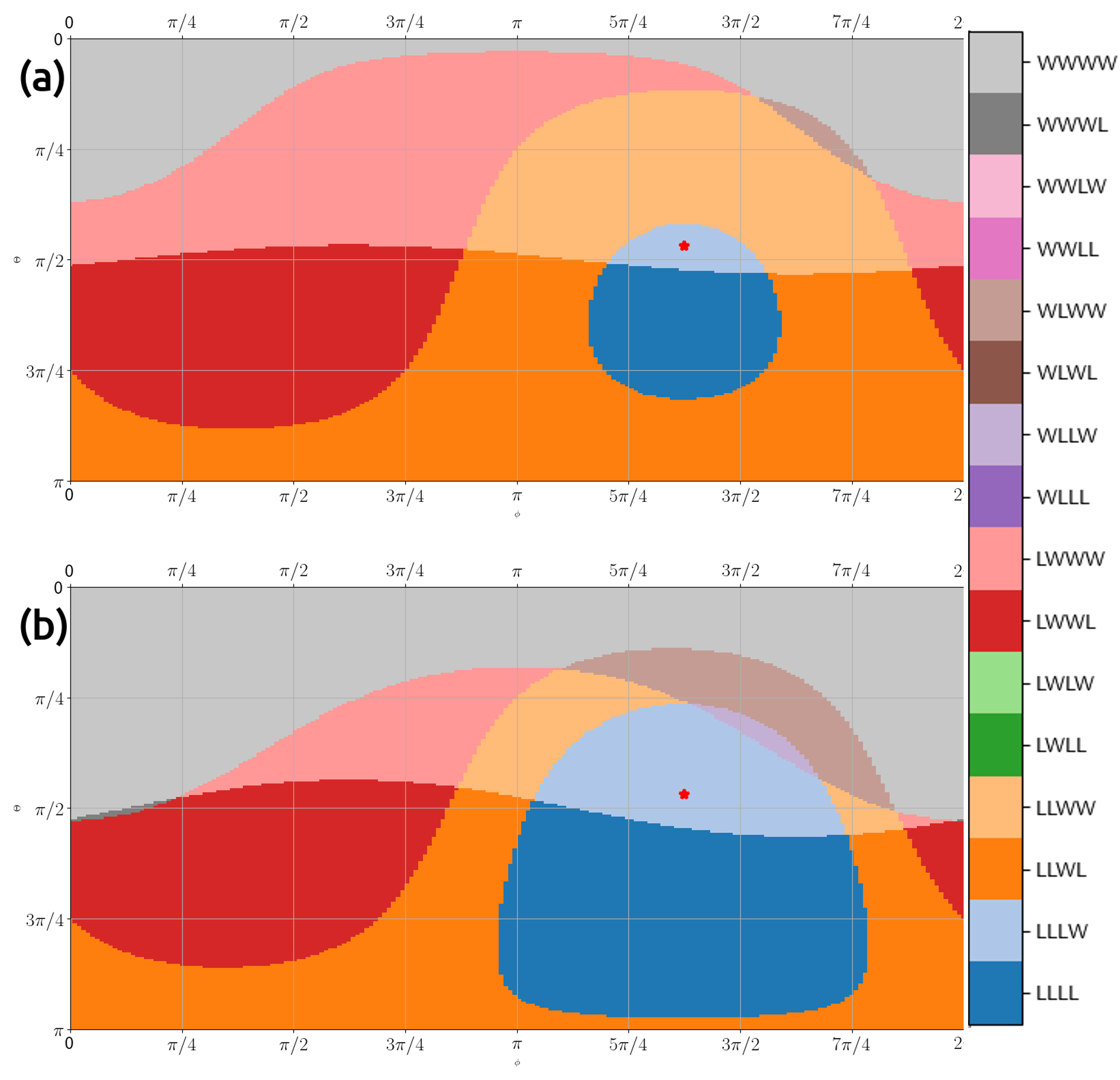}
	\caption{Identification of winning and losing regions on the Bloch sphere for the four walks of Eq. (\ref{eq:four_walks_PP}). Figs. (a) and (b) correspond to the operators $\hat{\mu}$ and $\hat{\Delta}$ respectively.  The home-state marked by a star in both the figures is $\stat{\boldsymbol{\Phi}}=0.741\stat{\boldsymbol{0}}-(0.257+0.62i)\stat{\boldsymbol{1}}$.}
	\label{fig:Four_walks_PP}
\end{figure}

\begin{figure*}[!ht]
	\centering
	\includegraphics[height=\textheight,width=\linewidth,keepaspectratio]{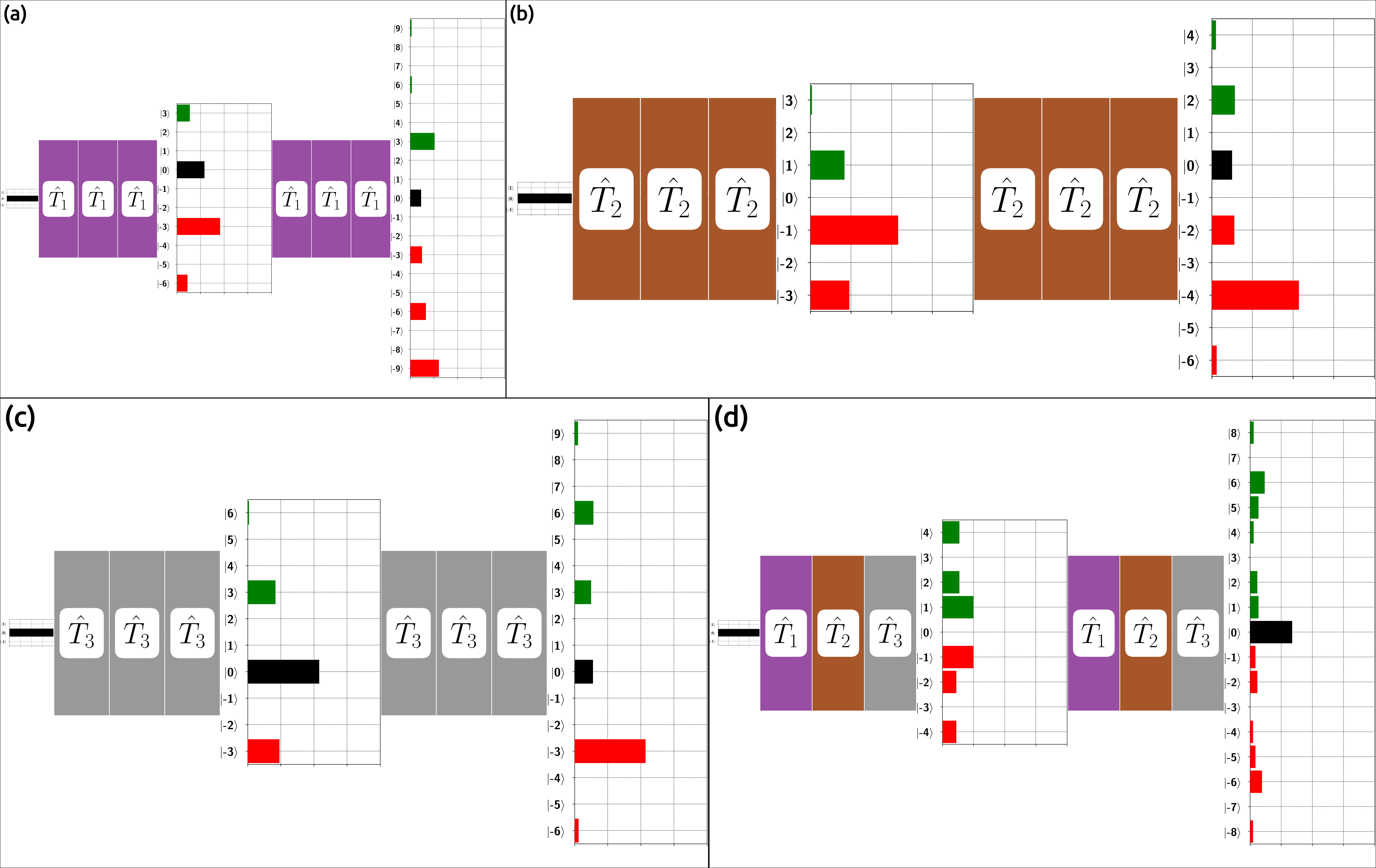}
	\caption{Histogram of obtained corresponding to the walks (a) $\hat{T}_1^{3n}$, (b) $\hat{T}_2^{3n}$ 
		(c) $\hat{T}_3^{3n}$ , and $[\hat{T}_3\hat{T}_2\hat{T}_1]^n$ for $n=1 \text{ and }2$, with the home-state $\stat{\boldsymbol{\Phi}}=0.741\stat{\boldsymbol{0}}-(0.257+0.62i)\stat{\boldsymbol{1}}$. The three steps $\hat{T}_1$, $\hat{T}_2$ and $\hat{T}_3$ are defined in Eq. (\ref{eq:Three_steps_PP}). }
	\label{fig:PP_3Step_Histogram}
\end{figure*}

Unlike the previous example of Eq. (\ref{eq:two_steps_PP}) where both the steps were unbiased, in this case two of the three steps are biased. \\
For this illustration we take the cycle length $n=2$, so that the four walks are:
\begin{equation}
	\begin{aligned}
		\hat{W}_1&=\hat{T_1}^6,\,\hat{W}_1=\hat{T_2}^6,\,\hat{W}_3=\hat{T_3}^6, \\
		\text{ and }\hat{W}_4&=\hat{T}_3\hat{T}_2\hat{T}_1\hat{T}_3\hat{T}_2\hat{T}_1.
	\end{aligned}
	\label{eq:four_walks_PP}
\end{equation}
Corresponding to these four walks, there are potentially sixteen combinations of winning or losing. These sixteen regions are identified on the Bloch sphere and is shown in Fig. (\ref{fig:Four_walks_PP}) as the $(\theta,\phi)$ plot. There are two images corresponding to the two operators $\hat{\mu}$ and $\hat{\Delta}$. In these, the desired Parrondo regions are the ones labeled ``LLLW''. 
Shown in Fig. (\ref{fig:PP_3Step_Histogram}).
The expectation values $\textbf{exp}(\hat{I}_2\otimes\hat{\mu},\hat{W}_i,\boldsymbol{\Phi})$, and $\textbf{exp}(\hat{I}_2\otimes\hat{\Delta},\hat{W}_i,\boldsymbol{\Phi})$ for the four walks $\hat{W}_i,i=1,\cdots,4$ with the home state $\stat{\boldsymbol{\Phi}}=0.741\stat{\boldsymbol{0}}-(0.257+0.62i)\stat{\boldsymbol{1}}$ are given in the table \ref{tab:PP_Three_step}. Evidently, the state $\stat{\boldsymbol{\Phi}}$ is a losing state for the three walks $\hat{W}_1,\,\hat{W}_2$, and $\hat{W}_3$ but is a winning state for the inhomogenous walk $\hat{W}_4$, for both the operators $\hat{\mu}$ and $\hat{\Delta}$.
\begin{table}
	\begin{tabular}{|c|c|c|c|c|}
		\hline
		&  $\hat{W}_1$&$\hat{W}_2$  &  $\hat{W}_3$ & $\hat{W}_4$ \\
		\hline
		$\hat{\mu}$&-3.402  &-2.204 & -0.306  & 0.334\\	\hline
		$\hat{\Delta}$&-0.334 &-0.535 & -0.271 & 0.08\\
		\hline
	\end{tabular}
	\caption{Parrondo paradox realized for the operators $\hat{\mu}$ and $\hat{\Delta}$, for the four walks of Eq. (\ref{eq:four_walks_PP}), corresponding to the three steps of Eq. (\ref{eq:Three_steps_PP}). }
	\label{tab:PP_Three_step}
\end{table}
\section{Conclusion}
\label{sec:Conclusions}
In discrete-time quantum walks, as a function of number of time-steps, the state of the walker evolves as a superposition of multiple terms of the coin-position composite space. As these quantum walks act, by design, in a translationally-invariant manner on the position-space, the properties of the resulting final states can be solely predicted by the initial coin-state.  Parrondo's paradox in this context refer to the phenomenon where, some initial coin-states are such that they are losing states for a walk comprising of two quantum steps which individually lead to losing states can collectively lead to a winning state when employed in a sequentially manner. Here winning states refer to those composite states in which the probability of occupying positive positions is greater than that of negative positions. \par
The aim of this paper has been to demonstrate the manifestation of Parrondo's paradox in a generalization of discrete-time quantum walks, on a novel winning criterion. In this work, a composite state is termed as a winning state if the expectation value of a given observable in that state happens to be greater than a certain cutoff value. 
Given a discrete time quantum walk and an observable as a Hermitian operator on the position-coin composite space, we have given an explicit procedure for identifying two orthogonal coin states that lead to the maximum and minimum expectation values of this observable. Furthermore, the expectation value corresponding to any other coin state can be obtained geometrically from these two coin states and the corresponding expectation values. With this, it is possible to explicitly identify coin states that lead to composite states wherein these expectation values are greater than a certain cutoff value. Armed with this, given a collection of quantum walks, and a Hermitian operator, we have given a scheme for identifying all the Parrondo states. These are coin-states which are losing states for all quantum walks composed of only one kinds of steps, but will be a winning state for a quantum walk comprised of applying all the given quantum steps in sequence. We have shown that these Parrondo states are not randomly scattered on the Bloch sphere, but occupy a contiguous convex region  on it. One could therefore identify a wedge of the Bloch sphere, every pure or mixed state of which is a Parrondo state. \par
Another signification contribution of this paper has been the introduction of a novel generalization of quantum walks. In these quantum walks, the steps could be biased, in the sense that forward and backward step-sizes can  be distinct. Indeed, a quantum step could be such that the walker move along the same direction but with different step-sizes, depending on the outcome of the coin-flip. All this, while being translationally-invariant. \par
This work also presented an innovative scheme for explicitly engineering such quantum steps so as to render a given coin state a Parrondo state. \par
We believe the novel quantum steps introduced here, and the procedure of identifying coin-state that lead to composite states having maximum expectation value of a given observable, would find applications extended beyond the realm of Parrondo paradox.

\FloatBarrier
\begin{acknowledgments}
	The author would like to thank Ms. Sridevi Ajikumar for useful discussion. 
\end{acknowledgments}
 \bibliography{PP}

\begin{thebibliography}{57}%
\makeatletter
\providecommand \@ifxundefined [1]{%
 \@ifx{#1\undefined}
}%
\providecommand \@ifnum [1]{%
 \ifnum #1\expandafter \@firstoftwo
 \else \expandafter \@secondoftwo
 \fi
}%
\providecommand \@ifx [1]{%
 \ifx #1\expandafter \@firstoftwo
 \else \expandafter \@secondoftwo
 \fi
}%
\providecommand \natexlab [1]{#1}%
\providecommand \enquote  [1]{``#1''}%
\providecommand \bibnamefont  [1]{#1}%
\providecommand \bibfnamefont [1]{#1}%
\providecommand \citenamefont [1]{#1}%
\providecommand \href@noop [0]{\@secondoftwo}%
\providecommand \href [0]{\begingroup \@sanitize@url \@href}%
\providecommand \@href[1]{\@@startlink{#1}\@@href}%
\providecommand \@@href[1]{\endgroup#1\@@endlink}%
\providecommand \@sanitize@url [0]{\catcode `\\12\catcode `\$12\catcode
  `\&12\catcode `\#12\catcode `\^12\catcode `\_12\catcode `\%12\relax}%
\providecommand \@@startlink[1]{}%
\providecommand \@@endlink[0]{}%
\providecommand \url  [0]{\begingroup\@sanitize@url \@url }%
\providecommand \@url [1]{\endgroup\@href {#1}{\urlprefix }}%
\providecommand \urlprefix  [0]{URL }%
\providecommand \Eprint [0]{\href }%
\providecommand \doibase [0]{https://doi.org/}%
\providecommand \selectlanguage [0]{\@gobble}%
\providecommand \bibinfo  [0]{\@secondoftwo}%
\providecommand \bibfield  [0]{\@secondoftwo}%
\providecommand \translation [1]{[#1]}%
\providecommand \BibitemOpen [0]{}%
\providecommand \bibitemStop [0]{}%
\providecommand \bibitemNoStop [0]{.\EOS\space}%
\providecommand \EOS [0]{\spacefactor3000\relax}%
\providecommand \BibitemShut  [1]{\csname bibitem#1\endcsname}%
\let\auto@bib@innerbib\@empty
\bibitem [{\citenamefont {Owen}(2013)}]{owen2013game}%
  \BibitemOpen
  \bibfield  {author} {\bibinfo {author} {\bibfnamefont {G.}~\bibnamefont
  {Owen}},\ }\href@noop {} {\emph {\bibinfo {title} {Game theory}}}\ (\bibinfo
  {publisher} {Emerald Group Publishing},\ \bibinfo {year} {2013})\BibitemShut
  {NoStop}%
\bibitem [{\citenamefont {Maschler}\ \emph {et~al.}(2020)\citenamefont
  {Maschler}, \citenamefont {Zamir},\ and\ \citenamefont
  {Solan}}]{maschler2020game}%
  \BibitemOpen
  \bibfield  {author} {\bibinfo {author} {\bibfnamefont {M.}~\bibnamefont
  {Maschler}}, \bibinfo {author} {\bibfnamefont {S.}~\bibnamefont {Zamir}},\
  and\ \bibinfo {author} {\bibfnamefont {E.}~\bibnamefont {Solan}},\
  }\href@noop {} {\emph {\bibinfo {title} {Game theory}}}\ (\bibinfo
  {publisher} {Cambridge University Press},\ \bibinfo {year}
  {2020})\BibitemShut {NoStop}%
\bibitem [{\citenamefont {Kolokoltsov}\ and\ \citenamefont
  {Malafeyev}(2020)}]{kolokoltsov2020understanding}%
  \BibitemOpen
  \bibfield  {author} {\bibinfo {author} {\bibfnamefont {V.~N.}\ \bibnamefont
  {Kolokoltsov}}\ and\ \bibinfo {author} {\bibfnamefont {O.~A.}\ \bibnamefont
  {Malafeyev}},\ }\href@noop {} {\emph {\bibinfo {title} {Understanding game
  theory: introduction to the analysis of many agent systems with competition
  and cooperation}}}\ (\bibinfo  {publisher} {World scientific},\ \bibinfo
  {year} {2020})\BibitemShut {NoStop}%
\bibitem [{\citenamefont {Parrondo}\ \emph {et~al.}(2000)\citenamefont
  {Parrondo}, \citenamefont {Harmer},\ and\ \citenamefont
  {Abbott}}]{parrondo2000new}%
  \BibitemOpen
  \bibfield  {author} {\bibinfo {author} {\bibfnamefont {J.~M.}\ \bibnamefont
  {Parrondo}}, \bibinfo {author} {\bibfnamefont {G.~P.}\ \bibnamefont
  {Harmer}},\ and\ \bibinfo {author} {\bibfnamefont {D.}~\bibnamefont
  {Abbott}},\ }\bibfield  {title} {\bibinfo {title} {New paradoxical games
  based on brownian ratchets},\ }\href@noop {} {\bibfield  {journal} {\bibinfo
  {journal} {Physical Review Letters}\ }\textbf {\bibinfo {volume} {85}},\
  \bibinfo {pages} {5226} (\bibinfo {year} {2000})}\BibitemShut {NoStop}%
\bibitem [{\citenamefont {Harmer}\ and\ \citenamefont
  {Abbott}(1999)}]{harmer1999losing}%
  \BibitemOpen
  \bibfield  {author} {\bibinfo {author} {\bibfnamefont {G.~P.}\ \bibnamefont
  {Harmer}}\ and\ \bibinfo {author} {\bibfnamefont {D.}~\bibnamefont
  {Abbott}},\ }\bibfield  {title} {\bibinfo {title} {Losing strategies can win
  by parrondo's paradox},\ }\href@noop {} {\bibfield  {journal} {\bibinfo
  {journal} {Nature}\ }\textbf {\bibinfo {volume} {402}},\ \bibinfo {pages}
  {864} (\bibinfo {year} {1999})}\BibitemShut {NoStop}%
\bibitem [{\citenamefont {Lai}\ and\ \citenamefont
  {Cheong}(2020{\natexlab{a}})}]{lai2020parrondo}%
  \BibitemOpen
  \bibfield  {author} {\bibinfo {author} {\bibfnamefont {J.~W.}\ \bibnamefont
  {Lai}}\ and\ \bibinfo {author} {\bibfnamefont {K.~H.}\ \bibnamefont
  {Cheong}},\ }\bibfield  {title} {\bibinfo {title} {Parrondo’s paradox from
  classical to quantum: A review},\ }\href@noop {} {\bibfield  {journal}
  {\bibinfo  {journal} {Nonlinear Dynamics}\ }\textbf {\bibinfo {volume}
  {100}},\ \bibinfo {pages} {849} (\bibinfo {year}
  {2020}{\natexlab{a}})}\BibitemShut {NoStop}%
\bibitem [{\citenamefont {Abbott}(2010)}]{abbott2010asymmetry}%
  \BibitemOpen
  \bibfield  {author} {\bibinfo {author} {\bibfnamefont {D.}~\bibnamefont
  {Abbott}},\ }\bibfield  {title} {\bibinfo {title} {Asymmetry and disorder: A
  decade of parrondo's paradox},\ }\href@noop {} {\bibfield  {journal}
  {\bibinfo  {journal} {Fluctuation and Noise Letters}\ }\textbf {\bibinfo
  {volume} {9}},\ \bibinfo {pages} {129} (\bibinfo {year} {2010})}\BibitemShut
  {NoStop}%
\bibitem [{\citenamefont {Canovas}\ and\ \citenamefont
  {Munoz}(2013)}]{canovas2013revisiting}%
  \BibitemOpen
  \bibfield  {author} {\bibinfo {author} {\bibfnamefont {J.~S.}\ \bibnamefont
  {Canovas}}\ and\ \bibinfo {author} {\bibfnamefont {M.}~\bibnamefont
  {Munoz}},\ }\bibfield  {title} {\bibinfo {title} {Revisiting parrondo's
  paradox for the logistic family},\ }\href@noop {} {\bibfield  {journal}
  {\bibinfo  {journal} {Fluctuation and Noise Letters}\ }\textbf {\bibinfo
  {volume} {12}},\ \bibinfo {pages} {1350015} (\bibinfo {year}
  {2013})}\BibitemShut {NoStop}%
\bibitem [{\citenamefont {Danca}\ \emph {et~al.}(2014)\citenamefont {Danca},
  \citenamefont {Fe{\v{c}}kan},\ and\ \citenamefont
  {Romera}}]{danca2014generalized}%
  \BibitemOpen
  \bibfield  {author} {\bibinfo {author} {\bibfnamefont {M.-F.}\ \bibnamefont
  {Danca}}, \bibinfo {author} {\bibfnamefont {M.}~\bibnamefont
  {Fe{\v{c}}kan}},\ and\ \bibinfo {author} {\bibfnamefont {M.}~\bibnamefont
  {Romera}},\ }\bibfield  {title} {\bibinfo {title} {Generalized form of
  parrondo's paradoxical game with applications to chaos control},\ }\href@noop
  {} {\bibfield  {journal} {\bibinfo  {journal} {International Journal of
  Bifurcation and Chaos}\ }\textbf {\bibinfo {volume} {24}},\ \bibinfo {pages}
  {1450008} (\bibinfo {year} {2014})}\BibitemShut {NoStop}%
\bibitem [{\citenamefont {Arena}\ \emph {et~al.}(2003)\citenamefont {Arena},
  \citenamefont {Fazzino}, \citenamefont {Fortuna},\ and\ \citenamefont
  {Maniscalco}}]{arena2003game}%
  \BibitemOpen
  \bibfield  {author} {\bibinfo {author} {\bibfnamefont {P.}~\bibnamefont
  {Arena}}, \bibinfo {author} {\bibfnamefont {S.}~\bibnamefont {Fazzino}},
  \bibinfo {author} {\bibfnamefont {L.}~\bibnamefont {Fortuna}},\ and\ \bibinfo
  {author} {\bibfnamefont {P.}~\bibnamefont {Maniscalco}},\ }\bibfield  {title}
  {\bibinfo {title} {Game theory and non-linear dynamics: the parrondo paradox
  case study},\ }\href@noop {} {\bibfield  {journal} {\bibinfo  {journal}
  {Chaos, Solitons \& Fractals}\ }\textbf {\bibinfo {volume} {17}},\ \bibinfo
  {pages} {545} (\bibinfo {year} {2003})}\BibitemShut {NoStop}%
\bibitem [{\citenamefont {Allison}\ and\ \citenamefont
  {Abbott}(2001)}]{allison2001control}%
  \BibitemOpen
  \bibfield  {author} {\bibinfo {author} {\bibfnamefont {A.}~\bibnamefont
  {Allison}}\ and\ \bibinfo {author} {\bibfnamefont {D.}~\bibnamefont
  {Abbott}},\ }\bibfield  {title} {\bibinfo {title} {Control systems with
  stochastic feedback},\ }\href@noop {} {\bibfield  {journal} {\bibinfo
  {journal} {Chaos: An Interdisciplinary Journal of Nonlinear Science}\
  }\textbf {\bibinfo {volume} {11}},\ \bibinfo {pages} {715} (\bibinfo {year}
  {2001})}\BibitemShut {NoStop}%
\bibitem [{\citenamefont {Lai}\ and\ \citenamefont
  {Cheong}(2020{\natexlab{b}})}]{lai2020social}%
  \BibitemOpen
  \bibfield  {author} {\bibinfo {author} {\bibfnamefont {J.~W.}\ \bibnamefont
  {Lai}}\ and\ \bibinfo {author} {\bibfnamefont {K.~H.}\ \bibnamefont
  {Cheong}},\ }\bibfield  {title} {\bibinfo {title} {Social dynamics and
  parrondo’s paradox: A narrative review},\ }\href@noop {} {\bibfield
  {journal} {\bibinfo  {journal} {Nonlinear Dynamics}\ }\textbf {\bibinfo
  {volume} {101}},\ \bibinfo {pages} {1} (\bibinfo {year}
  {2020}{\natexlab{b}})}\BibitemShut {NoStop}%
\bibitem [{\citenamefont {Lai}\ and\ \citenamefont
  {Cheong}(2024)}]{lai2024parrondo}%
  \BibitemOpen
  \bibfield  {author} {\bibinfo {author} {\bibfnamefont {J.~W.}\ \bibnamefont
  {Lai}}\ and\ \bibinfo {author} {\bibfnamefont {K.~H.}\ \bibnamefont
  {Cheong}},\ }\bibfield  {title} {\bibinfo {title} {A parrondo paradoxical
  interplay of reciprocity and reputation in social dynamics},\ }\href@noop {}
  {\bibfield  {journal} {\bibinfo  {journal} {Chaos, Solitons \& Fractals}\
  }\textbf {\bibinfo {volume} {179}},\ \bibinfo {pages} {114386} (\bibinfo
  {year} {2024})}\BibitemShut {NoStop}%
\bibitem [{\citenamefont {Cheong}\ \emph {et~al.}(2019)\citenamefont {Cheong},
  \citenamefont {Koh},\ and\ \citenamefont {Jones}}]{cheong2019paradoxical}%
  \BibitemOpen
  \bibfield  {author} {\bibinfo {author} {\bibfnamefont {K.~H.}\ \bibnamefont
  {Cheong}}, \bibinfo {author} {\bibfnamefont {J.~M.}\ \bibnamefont {Koh}},\
  and\ \bibinfo {author} {\bibfnamefont {M.~C.}\ \bibnamefont {Jones}},\
  }\bibfield  {title} {\bibinfo {title} {Paradoxical survival: examining the
  parrondo effect across biology},\ }\href@noop {} {\bibfield  {journal}
  {\bibinfo  {journal} {BioEssays}\ }\textbf {\bibinfo {volume} {41}},\
  \bibinfo {pages} {1900027} (\bibinfo {year} {2019})}\BibitemShut {NoStop}%
\bibitem [{\citenamefont {Cheong}\ \emph {et~al.}(2020)\citenamefont {Cheong},
  \citenamefont {Wen},\ and\ \citenamefont {Lai}}]{cheong2020relieving}%
  \BibitemOpen
  \bibfield  {author} {\bibinfo {author} {\bibfnamefont {K.~H.}\ \bibnamefont
  {Cheong}}, \bibinfo {author} {\bibfnamefont {T.}~\bibnamefont {Wen}},\ and\
  \bibinfo {author} {\bibfnamefont {J.~W.}\ \bibnamefont {Lai}},\ }\bibfield
  {title} {\bibinfo {title} {Relieving cost of epidemic by parrondo's paradox:
  a covid-19 case study},\ }\href@noop {} {\bibfield  {journal} {\bibinfo
  {journal} {Advanced Science}\ }\textbf {\bibinfo {volume} {7}},\ \bibinfo
  {pages} {2002324} (\bibinfo {year} {2020})}\BibitemShut {NoStop}%
\bibitem [{\citenamefont {Bassi}\ and\ \citenamefont
  {Ferrari}(2011)}]{bassi2011parrondos}%
  \BibitemOpen
  \bibfield  {author} {\bibinfo {author} {\bibfnamefont {C.}~\bibnamefont
  {Bassi}}\ and\ \bibinfo {author} {\bibfnamefont {P.}~\bibnamefont
  {Ferrari}},\ }\bibfield  {title} {\bibinfo {title} {Parrondo's paradox in
  financial markets},\ }\href@noop {} {\bibfield  {journal} {\bibinfo
  {journal} {Journal of Physics: Conference Series}\ }\textbf {\bibinfo
  {volume} {284}} (\bibinfo {year} {2011})}\BibitemShut {NoStop}%
\bibitem [{\citenamefont {Flitney}\ and\ \citenamefont
  {Abbott}(2002)}]{flitney2002introduction}%
  \BibitemOpen
  \bibfield  {author} {\bibinfo {author} {\bibfnamefont {A.~P.}\ \bibnamefont
  {Flitney}}\ and\ \bibinfo {author} {\bibfnamefont {D.}~\bibnamefont
  {Abbott}},\ }\bibfield  {title} {\bibinfo {title} {An introduction to quantum
  game theory},\ }\href@noop {} {\bibfield  {journal} {\bibinfo  {journal}
  {Fluctuation and Noise Letters}\ }\textbf {\bibinfo {volume} {2}},\ \bibinfo
  {pages} {R175} (\bibinfo {year} {2002})}\BibitemShut {NoStop}%
\bibitem [{\citenamefont {Khan}\ \emph {et~al.}(2018)\citenamefont {Khan},
  \citenamefont {Solmeyer}, \citenamefont {Balu},\ and\ \citenamefont
  {Humble}}]{khan2018quantum}%
  \BibitemOpen
  \bibfield  {author} {\bibinfo {author} {\bibfnamefont {F.~S.}\ \bibnamefont
  {Khan}}, \bibinfo {author} {\bibfnamefont {N.}~\bibnamefont {Solmeyer}},
  \bibinfo {author} {\bibfnamefont {R.}~\bibnamefont {Balu}},\ and\ \bibinfo
  {author} {\bibfnamefont {T.~S.}\ \bibnamefont {Humble}},\ }\bibfield  {title}
  {\bibinfo {title} {Quantum games: a review of the history, current state, and
  interpretation},\ }\href@noop {} {\bibfield  {journal} {\bibinfo  {journal}
  {Quantum Information Processing}\ }\textbf {\bibinfo {volume} {17}},\
  \bibinfo {pages} {1} (\bibinfo {year} {2018})}\BibitemShut {NoStop}%
\bibitem [{\citenamefont {Eisert}\ \emph {et~al.}(1999)\citenamefont {Eisert},
  \citenamefont {Wilkens},\ and\ \citenamefont
  {Lewenstein}}]{eisert1999quantum}%
  \BibitemOpen
  \bibfield  {author} {\bibinfo {author} {\bibfnamefont {J.}~\bibnamefont
  {Eisert}}, \bibinfo {author} {\bibfnamefont {M.}~\bibnamefont {Wilkens}},\
  and\ \bibinfo {author} {\bibfnamefont {M.}~\bibnamefont {Lewenstein}},\
  }\bibfield  {title} {\bibinfo {title} {Quantum games and quantum
  strategies},\ }\href@noop {} {\bibfield  {journal} {\bibinfo  {journal}
  {Physical Review Letters}\ }\textbf {\bibinfo {volume} {83}},\ \bibinfo
  {pages} {3077} (\bibinfo {year} {1999})}\BibitemShut {NoStop}%
\bibitem [{\citenamefont {Du}\ \emph {et~al.}(2002)\citenamefont {Du},
  \citenamefont {Li}, \citenamefont {Xu}, \citenamefont {Shi}, \citenamefont
  {Wu}, \citenamefont {Zhou},\ and\ \citenamefont {Han}}]{du2002experimental}%
  \BibitemOpen
  \bibfield  {author} {\bibinfo {author} {\bibfnamefont {J.}~\bibnamefont
  {Du}}, \bibinfo {author} {\bibfnamefont {H.}~\bibnamefont {Li}}, \bibinfo
  {author} {\bibfnamefont {X.}~\bibnamefont {Xu}}, \bibinfo {author}
  {\bibfnamefont {M.}~\bibnamefont {Shi}}, \bibinfo {author} {\bibfnamefont
  {J.}~\bibnamefont {Wu}}, \bibinfo {author} {\bibfnamefont {X.}~\bibnamefont
  {Zhou}},\ and\ \bibinfo {author} {\bibfnamefont {R.}~\bibnamefont {Han}},\
  }\bibfield  {title} {\bibinfo {title} {Experimental realization of quantum
  games on a quantum computer},\ }\href@noop {} {\bibfield  {journal} {\bibinfo
   {journal} {Physical Review Letters}\ }\textbf {\bibinfo {volume} {88}},\
  \bibinfo {pages} {137902} (\bibinfo {year} {2002})}\BibitemShut {NoStop}%
\bibitem [{\citenamefont {Portugal}(2013)}]{portugal2013quantum}%
  \BibitemOpen
  \bibfield  {author} {\bibinfo {author} {\bibfnamefont {R.}~\bibnamefont
  {Portugal}},\ }\href@noop {} {\emph {\bibinfo {title} {Quantum walks and
  search algorithms}}},\ Vol.~\bibinfo {volume} {19}\ (\bibinfo  {publisher}
  {Springer},\ \bibinfo {year} {2013})\BibitemShut {NoStop}%
\bibitem [{\citenamefont {Mackay}\ \emph {et~al.}(2002)\citenamefont {Mackay},
  \citenamefont {Bartlett}, \citenamefont {Stephenson},\ and\ \citenamefont
  {Sanders}}]{mackay2002quantum}%
  \BibitemOpen
  \bibfield  {author} {\bibinfo {author} {\bibfnamefont {T.~D.}\ \bibnamefont
  {Mackay}}, \bibinfo {author} {\bibfnamefont {S.~D.}\ \bibnamefont
  {Bartlett}}, \bibinfo {author} {\bibfnamefont {L.~T.}\ \bibnamefont
  {Stephenson}},\ and\ \bibinfo {author} {\bibfnamefont {B.~C.}\ \bibnamefont
  {Sanders}},\ }\bibfield  {title} {\bibinfo {title} {Quantum walks in higher
  dimensions},\ }\href@noop {} {\bibfield  {journal} {\bibinfo  {journal}
  {Journal of Physics A: Mathematical and General}\ }\textbf {\bibinfo {volume}
  {35}},\ \bibinfo {pages} {2745} (\bibinfo {year} {2002})}\BibitemShut
  {NoStop}%
\bibitem [{\citenamefont {Venegas-Andraca}(2012)}]{VenegasAndraca2012}%
  \BibitemOpen
  \bibfield  {author} {\bibinfo {author} {\bibfnamefont {S.~E.}\ \bibnamefont
  {Venegas-Andraca}},\ }\bibfield  {title} {\bibinfo {title} {Quantum walks: a
  comprehensive review},\ }\href@noop {} {\bibfield  {journal} {\bibinfo
  {journal} {Quantum Information Processing}\ }\textbf {\bibinfo {volume}
  {11}},\ \bibinfo {pages} {1015} (\bibinfo {year} {2012})}\BibitemShut
  {NoStop}%
\bibitem [{\citenamefont {Lovett}\ \emph {et~al.}(2010)\citenamefont {Lovett},
  \citenamefont {Cooper}, \citenamefont {Everitt}, \citenamefont {Trevers},\
  and\ \citenamefont {Kendon}}]{lovett2010universal}%
  \BibitemOpen
  \bibfield  {author} {\bibinfo {author} {\bibfnamefont {N.~B.}\ \bibnamefont
  {Lovett}}, \bibinfo {author} {\bibfnamefont {S.}~\bibnamefont {Cooper}},
  \bibinfo {author} {\bibfnamefont {M.}~\bibnamefont {Everitt}}, \bibinfo
  {author} {\bibfnamefont {M.}~\bibnamefont {Trevers}},\ and\ \bibinfo {author}
  {\bibfnamefont {V.}~\bibnamefont {Kendon}},\ }\bibfield  {title} {\bibinfo
  {title} {Universal quantum computation using the discrete-time quantum
  walk},\ }\href@noop {} {\bibfield  {journal} {\bibinfo  {journal} {Physical
  Review A}\ }\textbf {\bibinfo {volume} {81}},\ \bibinfo {pages} {042330}
  (\bibinfo {year} {2010})}\BibitemShut {NoStop}%
\bibitem [{\citenamefont {Singh}\ \emph {et~al.}(2021)\citenamefont {Singh},
  \citenamefont {Chawla}, \citenamefont {Sarkar},\ and\ \citenamefont
  {Chandrashekar}}]{singh2021universal}%
  \BibitemOpen
  \bibfield  {author} {\bibinfo {author} {\bibfnamefont {S.}~\bibnamefont
  {Singh}}, \bibinfo {author} {\bibfnamefont {P.}~\bibnamefont {Chawla}},
  \bibinfo {author} {\bibfnamefont {A.}~\bibnamefont {Sarkar}},\ and\ \bibinfo
  {author} {\bibfnamefont {C.}~\bibnamefont {Chandrashekar}},\ }\bibfield
  {title} {\bibinfo {title} {Universal quantum computing using single-particle
  discrete-time quantum walk},\ }\href@noop {} {\bibfield  {journal} {\bibinfo
  {journal} {Scientific Reports}\ }\textbf {\bibinfo {volume} {11}},\ \bibinfo
  {pages} {11551} (\bibinfo {year} {2021})}\BibitemShut {NoStop}%
\bibitem [{\citenamefont {Asaka}\ \emph {et~al.}(2023)\citenamefont {Asaka},
  \citenamefont {Sakai},\ and\ \citenamefont {Yahagi}}]{asaka2023two}%
  \BibitemOpen
  \bibfield  {author} {\bibinfo {author} {\bibfnamefont {R.}~\bibnamefont
  {Asaka}}, \bibinfo {author} {\bibfnamefont {K.}~\bibnamefont {Sakai}},\ and\
  \bibinfo {author} {\bibfnamefont {R.}~\bibnamefont {Yahagi}},\ }\bibfield
  {title} {\bibinfo {title} {Two-level quantum walkers on directed graphs. i.
  universal quantum computing},\ }\href@noop {} {\bibfield  {journal} {\bibinfo
   {journal} {Physical Review A}\ }\textbf {\bibinfo {volume} {107}},\ \bibinfo
  {pages} {022415} (\bibinfo {year} {2023})}\BibitemShut {NoStop}%
\bibitem [{\citenamefont {Kendon}(2006)}]{kendon2006random}%
  \BibitemOpen
  \bibfield  {author} {\bibinfo {author} {\bibfnamefont {V.~M.}\ \bibnamefont
  {Kendon}},\ }\bibfield  {title} {\bibinfo {title} {A random walk approach to
  quantum algorithms},\ }\href@noop {} {\bibfield  {journal} {\bibinfo
  {journal} {Philosophical Transactions of the Royal Society A: Mathematical,
  Physical and Engineering Sciences}\ }\textbf {\bibinfo {volume} {364}},\
  \bibinfo {pages} {3407} (\bibinfo {year} {2006})}\BibitemShut {NoStop}%
\bibitem [{\citenamefont {Childs}\ \emph {et~al.}(2003)\citenamefont {Childs},
  \citenamefont {Cleve}, \citenamefont {Deotto}, \citenamefont {Farhi},
  \citenamefont {Gutmann},\ and\ \citenamefont
  {Spielman}}]{childs2003exponential}%
  \BibitemOpen
  \bibfield  {author} {\bibinfo {author} {\bibfnamefont {A.~M.}\ \bibnamefont
  {Childs}}, \bibinfo {author} {\bibfnamefont {R.}~\bibnamefont {Cleve}},
  \bibinfo {author} {\bibfnamefont {E.}~\bibnamefont {Deotto}}, \bibinfo
  {author} {\bibfnamefont {E.}~\bibnamefont {Farhi}}, \bibinfo {author}
  {\bibfnamefont {S.}~\bibnamefont {Gutmann}},\ and\ \bibinfo {author}
  {\bibfnamefont {D.~A.}\ \bibnamefont {Spielman}},\ }\bibfield  {title}
  {\bibinfo {title} {Exponential algorithmic speedup by a quantum walk},\ }in\
  \href@noop {} {\emph {\bibinfo {booktitle} {Proceedings of the thirty-fifth
  annual ACM symposium on Theory of computing}}}\ (\bibinfo {year} {2003})\
  pp.\ \bibinfo {pages} {59--68}\BibitemShut {NoStop}%
\bibitem [{\citenamefont {Apers}\ \emph {et~al.}(2022)\citenamefont {Apers},
  \citenamefont {Chakraborty}, \citenamefont {Novo},\ and\ \citenamefont
  {Roland}}]{apers2022quadratic}%
  \BibitemOpen
  \bibfield  {author} {\bibinfo {author} {\bibfnamefont {S.}~\bibnamefont
  {Apers}}, \bibinfo {author} {\bibfnamefont {S.}~\bibnamefont {Chakraborty}},
  \bibinfo {author} {\bibfnamefont {L.}~\bibnamefont {Novo}},\ and\ \bibinfo
  {author} {\bibfnamefont {J.}~\bibnamefont {Roland}},\ }\bibfield  {title}
  {\bibinfo {title} {Quadratic speedup for spatial search by continuous-time
  quantum walk},\ }\href@noop {} {\bibfield  {journal} {\bibinfo  {journal}
  {Physical review letters}\ }\textbf {\bibinfo {volume} {129}},\ \bibinfo
  {pages} {160502} (\bibinfo {year} {2022})}\BibitemShut {NoStop}%
\bibitem [{\citenamefont {Lovett}\ \emph {et~al.}(2019)\citenamefont {Lovett},
  \citenamefont {Everitt}, \citenamefont {Heath},\ and\ \citenamefont
  {Kendon}}]{lovett2019quantum}%
  \BibitemOpen
  \bibfield  {author} {\bibinfo {author} {\bibfnamefont {N.~B.}\ \bibnamefont
  {Lovett}}, \bibinfo {author} {\bibfnamefont {M.}~\bibnamefont {Everitt}},
  \bibinfo {author} {\bibfnamefont {R.~M.}\ \bibnamefont {Heath}},\ and\
  \bibinfo {author} {\bibfnamefont {V.}~\bibnamefont {Kendon}},\ }\bibfield
  {title} {\bibinfo {title} {The quantum walk search algorithm: Factors
  affecting efficiency},\ }\href@noop {} {\bibfield  {journal} {\bibinfo
  {journal} {Mathematical Structures in Computer Science}\ }\textbf {\bibinfo
  {volume} {29}},\ \bibinfo {pages} {389} (\bibinfo {year} {2019})}\BibitemShut
  {NoStop}%
\bibitem [{\citenamefont {Zhou}\ \emph {et~al.}(2019)\citenamefont {Zhou},
  \citenamefont {Cai}, \citenamefont {Su},\ and\ \citenamefont
  {Yang}}]{zhou2019protocol}%
  \BibitemOpen
  \bibfield  {author} {\bibinfo {author} {\bibfnamefont {J.-Q.}\ \bibnamefont
  {Zhou}}, \bibinfo {author} {\bibfnamefont {L.}~\bibnamefont {Cai}}, \bibinfo
  {author} {\bibfnamefont {Q.-P.}\ \bibnamefont {Su}},\ and\ \bibinfo {author}
  {\bibfnamefont {C.-P.}\ \bibnamefont {Yang}},\ }\bibfield  {title} {\bibinfo
  {title} {Protocol of a quantum walk in circuit qed},\ }\href@noop {}
  {\bibfield  {journal} {\bibinfo  {journal} {Physical Review A}\ }\textbf
  {\bibinfo {volume} {100}},\ \bibinfo {pages} {012343} (\bibinfo {year}
  {2019})}\BibitemShut {NoStop}%
\bibitem [{\citenamefont {Su}\ \emph {et~al.}(2019)\citenamefont {Su},
  \citenamefont {Zhang}, \citenamefont {Yu}, \citenamefont {Zhou},
  \citenamefont {Jin}, \citenamefont {Xu}, \citenamefont {Xiong}, \citenamefont
  {Xu}, \citenamefont {Sun}, \citenamefont {Chen} \emph
  {et~al.}}]{su2019experimental}%
  \BibitemOpen
  \bibfield  {author} {\bibinfo {author} {\bibfnamefont {Q.-P.}\ \bibnamefont
  {Su}}, \bibinfo {author} {\bibfnamefont {Y.}~\bibnamefont {Zhang}}, \bibinfo
  {author} {\bibfnamefont {L.}~\bibnamefont {Yu}}, \bibinfo {author}
  {\bibfnamefont {J.-Q.}\ \bibnamefont {Zhou}}, \bibinfo {author}
  {\bibfnamefont {J.-S.}\ \bibnamefont {Jin}}, \bibinfo {author} {\bibfnamefont
  {X.-Q.}\ \bibnamefont {Xu}}, \bibinfo {author} {\bibfnamefont {S.-J.}\
  \bibnamefont {Xiong}}, \bibinfo {author} {\bibfnamefont {Q.}~\bibnamefont
  {Xu}}, \bibinfo {author} {\bibfnamefont {Z.}~\bibnamefont {Sun}}, \bibinfo
  {author} {\bibfnamefont {K.}~\bibnamefont {Chen}}, \emph {et~al.},\
  }\bibfield  {title} {\bibinfo {title} {Experimental demonstration of quantum
  walks with initial superposition states},\ }\href@noop {} {\bibfield
  {journal} {\bibinfo  {journal} {npj Quantum Information}\ }\textbf {\bibinfo
  {volume} {5}},\ \bibinfo {pages} {40} (\bibinfo {year} {2019})}\BibitemShut
  {NoStop}%
\bibitem [{\citenamefont {Giordani}\ \emph {et~al.}(2019)\citenamefont
  {Giordani}, \citenamefont {Polino}, \citenamefont {Emiliani}, \citenamefont
  {Suprano}, \citenamefont {Innocenti}, \citenamefont {Majury}, \citenamefont
  {Marrucci}, \citenamefont {Paternostro}, \citenamefont {Ferraro},
  \citenamefont {Spagnolo} \emph {et~al.}}]{giordani2019experimental}%
  \BibitemOpen
  \bibfield  {author} {\bibinfo {author} {\bibfnamefont {T.}~\bibnamefont
  {Giordani}}, \bibinfo {author} {\bibfnamefont {E.}~\bibnamefont {Polino}},
  \bibinfo {author} {\bibfnamefont {S.}~\bibnamefont {Emiliani}}, \bibinfo
  {author} {\bibfnamefont {A.}~\bibnamefont {Suprano}}, \bibinfo {author}
  {\bibfnamefont {L.}~\bibnamefont {Innocenti}}, \bibinfo {author}
  {\bibfnamefont {H.}~\bibnamefont {Majury}}, \bibinfo {author} {\bibfnamefont
  {L.}~\bibnamefont {Marrucci}}, \bibinfo {author} {\bibfnamefont
  {M.}~\bibnamefont {Paternostro}}, \bibinfo {author} {\bibfnamefont
  {A.}~\bibnamefont {Ferraro}}, \bibinfo {author} {\bibfnamefont
  {N.}~\bibnamefont {Spagnolo}}, \emph {et~al.},\ }\bibfield  {title} {\bibinfo
  {title} {Experimental engineering of arbitrary qudit states with
  discrete-time quantum walks},\ }\href@noop {} {\bibfield  {journal} {\bibinfo
   {journal} {Physical review letters}\ }\textbf {\bibinfo {volume} {122}},\
  \bibinfo {pages} {020503} (\bibinfo {year} {2019})}\BibitemShut {NoStop}%
\bibitem [{\citenamefont {Matjeschk}\ \emph {et~al.}(2012)\citenamefont
  {Matjeschk}, \citenamefont {Schneider}, \citenamefont {Enderlein},
  \citenamefont {Huber}, \citenamefont {Schmitz}, \citenamefont {Glueckert},\
  and\ \citenamefont {Schaetz}}]{matjeschk2012experimental}%
  \BibitemOpen
  \bibfield  {author} {\bibinfo {author} {\bibfnamefont {R.}~\bibnamefont
  {Matjeschk}}, \bibinfo {author} {\bibfnamefont {C.}~\bibnamefont
  {Schneider}}, \bibinfo {author} {\bibfnamefont {M.}~\bibnamefont
  {Enderlein}}, \bibinfo {author} {\bibfnamefont {T.}~\bibnamefont {Huber}},
  \bibinfo {author} {\bibfnamefont {H.}~\bibnamefont {Schmitz}}, \bibinfo
  {author} {\bibfnamefont {J.}~\bibnamefont {Glueckert}},\ and\ \bibinfo
  {author} {\bibfnamefont {T.}~\bibnamefont {Schaetz}},\ }\bibfield  {title}
  {\bibinfo {title} {Experimental simulation and limitations of quantum walks
  with trapped ions},\ }\href@noop {} {\bibfield  {journal} {\bibinfo
  {journal} {New Journal of Physics}\ }\textbf {\bibinfo {volume} {14}},\
  \bibinfo {pages} {035012} (\bibinfo {year} {2012})}\BibitemShut {NoStop}%
\bibitem [{\citenamefont {Flitney}\ \emph {et~al.}(2004)\citenamefont
  {Flitney}, \citenamefont {Abbott},\ and\ \citenamefont
  {Johnson}}]{flitney2004quantum}%
  \BibitemOpen
  \bibfield  {author} {\bibinfo {author} {\bibfnamefont {A.~P.}\ \bibnamefont
  {Flitney}}, \bibinfo {author} {\bibfnamefont {D.}~\bibnamefont {Abbott}},\
  and\ \bibinfo {author} {\bibfnamefont {N.~F.}\ \bibnamefont {Johnson}},\
  }\bibfield  {title} {\bibinfo {title} {Quantum walks with history
  dependence},\ }\href@noop {} {\bibfield  {journal} {\bibinfo  {journal}
  {Journal of Physics A: Mathematical and General}\ }\textbf {\bibinfo {volume}
  {37}},\ \bibinfo {pages} {7581} (\bibinfo {year} {2004})}\BibitemShut
  {NoStop}%
\bibitem [{\citenamefont {Walczak}\ and\ \citenamefont
  {Bauer}(2022)}]{walczak2022parrondothreecoins}%
  \BibitemOpen
  \bibfield  {author} {\bibinfo {author} {\bibfnamefont {Z.}~\bibnamefont
  {Walczak}}\ and\ \bibinfo {author} {\bibfnamefont {J.~H.}\ \bibnamefont
  {Bauer}},\ }\bibfield  {title} {\bibinfo {title} {Parrondo's paradox in
  quantum walks with three coins},\ }\href@noop {} {\bibfield  {journal}
  {\bibinfo  {journal} {Physical Review E}\ }\textbf {\bibinfo {volume}
  {105}},\ \bibinfo {pages} {064211} (\bibinfo {year} {2022})}\BibitemShut
  {NoStop}%
\bibitem [{\citenamefont {Lai}\ \emph {et~al.}(2020)\citenamefont {Lai},
  \citenamefont {Tan}, \citenamefont {Lu}, \citenamefont {Yap},\ and\
  \citenamefont {Cheong}}]{lai2020parrondofoursided}%
  \BibitemOpen
  \bibfield  {author} {\bibinfo {author} {\bibfnamefont {J.~W.}\ \bibnamefont
  {Lai}}, \bibinfo {author} {\bibfnamefont {J.~R.~A.}\ \bibnamefont {Tan}},
  \bibinfo {author} {\bibfnamefont {H.}~\bibnamefont {Lu}}, \bibinfo {author}
  {\bibfnamefont {Z.~R.}\ \bibnamefont {Yap}},\ and\ \bibinfo {author}
  {\bibfnamefont {K.~H.}\ \bibnamefont {Cheong}},\ }\bibfield  {title}
  {\bibinfo {title} {Parrondo paradoxical walk using four-sided quantum
  coins},\ }\href@noop {} {\bibfield  {journal} {\bibinfo  {journal} {Physical
  Review E}\ }\textbf {\bibinfo {volume} {102}},\ \bibinfo {pages} {012213}
  (\bibinfo {year} {2020})}\BibitemShut {NoStop}%
\bibitem [{\citenamefont {Pires}\ and\ \citenamefont
  {Queir{\'o}s}(2020)}]{pires2020parrondo}%
  \BibitemOpen
  \bibfield  {author} {\bibinfo {author} {\bibfnamefont {M.~A.}\ \bibnamefont
  {Pires}}\ and\ \bibinfo {author} {\bibfnamefont {S.~M.~D.}\ \bibnamefont
  {Queir{\'o}s}},\ }\bibfield  {title} {\bibinfo {title} {Parrondo's paradox in
  quantum walks with time-dependent coin operators},\ }\href@noop {} {\bibfield
   {journal} {\bibinfo  {journal} {Physical Review E}\ }\textbf {\bibinfo
  {volume} {102}},\ \bibinfo {pages} {042124} (\bibinfo {year}
  {2020})}\BibitemShut {NoStop}%
\bibitem [{\citenamefont {Walczak}\ and\ \citenamefont
  {Bauer}(2021)}]{walczak2021parrondo}%
  \BibitemOpen
  \bibfield  {author} {\bibinfo {author} {\bibfnamefont {Z.}~\bibnamefont
  {Walczak}}\ and\ \bibinfo {author} {\bibfnamefont {J.~H.}\ \bibnamefont
  {Bauer}},\ }\bibfield  {title} {\bibinfo {title} {Parrondo's paradox in
  quantum walks with deterministic aperiodic sequence of coins},\ }\href@noop
  {} {\bibfield  {journal} {\bibinfo  {journal} {Physical Review E}\ }\textbf
  {\bibinfo {volume} {104}},\ \bibinfo {pages} {064209} (\bibinfo {year}
  {2021})}\BibitemShut {NoStop}%
\bibitem [{\citenamefont {Trautmann}\ \emph {et~al.}(2022)\citenamefont
  {Trautmann}, \citenamefont {Groiseau},\ and\ \citenamefont
  {Wimberger}}]{trautmann2022parrondo}%
  \BibitemOpen
  \bibfield  {author} {\bibinfo {author} {\bibfnamefont {G.}~\bibnamefont
  {Trautmann}}, \bibinfo {author} {\bibfnamefont {C.}~\bibnamefont
  {Groiseau}},\ and\ \bibinfo {author} {\bibfnamefont {S.}~\bibnamefont
  {Wimberger}},\ }\bibfield  {title} {\bibinfo {title} {Parrondo’s paradox
  for discrete-time quantum walks in momentum space},\ }\href@noop {}
  {\bibfield  {journal} {\bibinfo  {journal} {Fluctuation and Noise Letters}\
  }\textbf {\bibinfo {volume} {21}},\ \bibinfo {pages} {2250053} (\bibinfo
  {year} {2022})}\BibitemShut {NoStop}%
\bibitem [{\citenamefont {Panda}\ \emph {et~al.}(2022)\citenamefont {Panda},
  \citenamefont {Govind},\ and\ \citenamefont
  {Benjamin}}]{panda2022generating}%
  \BibitemOpen
  \bibfield  {author} {\bibinfo {author} {\bibfnamefont {D.~K.}\ \bibnamefont
  {Panda}}, \bibinfo {author} {\bibfnamefont {B.~V.}\ \bibnamefont {Govind}},\
  and\ \bibinfo {author} {\bibfnamefont {C.}~\bibnamefont {Benjamin}},\
  }\bibfield  {title} {\bibinfo {title} {Generating highly entangled states via
  discrete-time quantum walks with parrondo sequences},\ }\href@noop {}
  {\bibfield  {journal} {\bibinfo  {journal} {Physica A: Statistical Mechanics
  and its Applications}\ }\textbf {\bibinfo {volume} {608}},\ \bibinfo {pages}
  {128256} (\bibinfo {year} {2022})}\BibitemShut {NoStop}%
\bibitem [{\citenamefont {Fang}\ \emph {et~al.}(2023)\citenamefont {Fang},
  \citenamefont {An}, \citenamefont {Zhang}, \citenamefont {Sanders},\ and\
  \citenamefont {Lu}}]{fang2023maximal}%
  \BibitemOpen
  \bibfield  {author} {\bibinfo {author} {\bibfnamefont {X.-X.}\ \bibnamefont
  {Fang}}, \bibinfo {author} {\bibfnamefont {K.}~\bibnamefont {An}}, \bibinfo
  {author} {\bibfnamefont {B.-T.}\ \bibnamefont {Zhang}}, \bibinfo {author}
  {\bibfnamefont {B.~C.}\ \bibnamefont {Sanders}},\ and\ \bibinfo {author}
  {\bibfnamefont {H.}~\bibnamefont {Lu}},\ }\bibfield  {title} {\bibinfo
  {title} {Maximal coin-position entanglement generation in a quantum walk for
  the third step and beyond regardless of the initial state},\ }\href@noop {}
  {\bibfield  {journal} {\bibinfo  {journal} {Physical Review A}\ }\textbf
  {\bibinfo {volume} {107}},\ \bibinfo {pages} {012433} (\bibinfo {year}
  {2023})}\BibitemShut {NoStop}%
\bibitem [{\citenamefont {Walczak}\ and\ \citenamefont
  {Bauer}(2023)}]{walczak2023noise}%
  \BibitemOpen
  \bibfield  {author} {\bibinfo {author} {\bibfnamefont {Z.}~\bibnamefont
  {Walczak}}\ and\ \bibinfo {author} {\bibfnamefont {J.~H.}\ \bibnamefont
  {Bauer}},\ }\bibfield  {title} {\bibinfo {title} {Noise-induced parrondo's
  paradox in discrete-time quantum walks},\ }\href@noop {} {\bibfield
  {journal} {\bibinfo  {journal} {Physical Review E}\ }\textbf {\bibinfo
  {volume} {108}},\ \bibinfo {pages} {044212} (\bibinfo {year}
  {2023})}\BibitemShut {NoStop}%
\bibitem [{\citenamefont {Lai}\ and\ \citenamefont
  {Cheong}(2021)}]{lai2021chaotic}%
  \BibitemOpen
  \bibfield  {author} {\bibinfo {author} {\bibfnamefont {J.~W.}\ \bibnamefont
  {Lai}}\ and\ \bibinfo {author} {\bibfnamefont {K.~H.}\ \bibnamefont
  {Cheong}},\ }\bibfield  {title} {\bibinfo {title} {Chaotic switching for
  quantum coin parrondo's games with application to encryption},\ }\href@noop
  {} {\bibfield  {journal} {\bibinfo  {journal} {Physical Review Research}\
  }\textbf {\bibinfo {volume} {3}},\ \bibinfo {pages} {L022019} (\bibinfo
  {year} {2021})}\BibitemShut {NoStop}%
\bibitem [{\citenamefont {Jan}\ \emph {et~al.}(2020)\citenamefont {Jan},
  \citenamefont {Wang}, \citenamefont {Xu}, \citenamefont {Pan}, \citenamefont
  {Chen}, \citenamefont {Han}, \citenamefont {Li}, \citenamefont {Guo},\ and\
  \citenamefont {Abbott}}]{jan2020experimental}%
  \BibitemOpen
  \bibfield  {author} {\bibinfo {author} {\bibfnamefont {M.}~\bibnamefont
  {Jan}}, \bibinfo {author} {\bibfnamefont {Q.-Q.}\ \bibnamefont {Wang}},
  \bibinfo {author} {\bibfnamefont {X.-Y.}\ \bibnamefont {Xu}}, \bibinfo
  {author} {\bibfnamefont {W.-W.}\ \bibnamefont {Pan}}, \bibinfo {author}
  {\bibfnamefont {Z.}~\bibnamefont {Chen}}, \bibinfo {author} {\bibfnamefont
  {Y.-J.}\ \bibnamefont {Han}}, \bibinfo {author} {\bibfnamefont {C.-F.}\
  \bibnamefont {Li}}, \bibinfo {author} {\bibfnamefont {G.-C.}\ \bibnamefont
  {Guo}},\ and\ \bibinfo {author} {\bibfnamefont {D.}~\bibnamefont {Abbott}},\
  }\bibfield  {title} {\bibinfo {title} {Experimental realization of parrondo's
  paradox in 1d quantum walks},\ }\href@noop {} {\bibfield  {journal} {\bibinfo
   {journal} {Advanced Quantum Technologies}\ }\textbf {\bibinfo {volume}
  {3}},\ \bibinfo {pages} {1900127} (\bibinfo {year} {2020})}\BibitemShut
  {NoStop}%
\bibitem [{\citenamefont {Lai}\ and\ \citenamefont
  {Cheong}(2020{\natexlab{c}})}]{lai2020parrondoPRE}%
  \BibitemOpen
  \bibfield  {author} {\bibinfo {author} {\bibfnamefont {J.~W.}\ \bibnamefont
  {Lai}}\ and\ \bibinfo {author} {\bibfnamefont {K.~H.}\ \bibnamefont
  {Cheong}},\ }\bibfield  {title} {\bibinfo {title} {Parrondo effect in quantum
  coin-toss simulations},\ }\href@noop {} {\bibfield  {journal} {\bibinfo
  {journal} {Physical review E}\ }\textbf {\bibinfo {volume} {101}},\ \bibinfo
  {pages} {052212} (\bibinfo {year} {2020}{\natexlab{c}})}\BibitemShut
  {NoStop}%
\bibitem [{\citenamefont {Jan}\ \emph {et~al.}(2023)\citenamefont {Jan},
  \citenamefont {Khan},\ and\ \citenamefont {Xianlong}}]{jan2023territories}%
  \BibitemOpen
  \bibfield  {author} {\bibinfo {author} {\bibfnamefont {M.}~\bibnamefont
  {Jan}}, \bibinfo {author} {\bibfnamefont {N.~A.}\ \bibnamefont {Khan}},\ and\
  \bibinfo {author} {\bibfnamefont {G.}~\bibnamefont {Xianlong}},\ }\bibfield
  {title} {\bibinfo {title} {Territories of parrondo’s paradox and its
  entanglement dynamics in quantum walks},\ }\href@noop {} {\bibfield
  {journal} {\bibinfo  {journal} {The European Physical Journal Plus}\ }\textbf
  {\bibinfo {volume} {138}},\ \bibinfo {pages} {65} (\bibinfo {year}
  {2023})}\BibitemShut {NoStop}%
\bibitem [{\citenamefont {Chandrashekar}\ and\ \citenamefont
  {Banerjee}(2011)}]{chandrashekar2011parrondo}%
  \BibitemOpen
  \bibfield  {author} {\bibinfo {author} {\bibfnamefont {C.~M.}\ \bibnamefont
  {Chandrashekar}}\ and\ \bibinfo {author} {\bibfnamefont {S.}~\bibnamefont
  {Banerjee}},\ }\bibfield  {title} {\bibinfo {title} {Parrondo's game using a
  discrete-time quantum walk},\ }\href@noop {} {\bibfield  {journal} {\bibinfo
  {journal} {Physics Letters A}\ }\textbf {\bibinfo {volume} {375}},\ \bibinfo
  {pages} {1553} (\bibinfo {year} {2011})}\BibitemShut {NoStop}%
\bibitem [{\citenamefont {Rajendran}\ and\ \citenamefont
  {Benjamin}(2018)}]{rajendran2018implementing}%
  \BibitemOpen
  \bibfield  {author} {\bibinfo {author} {\bibfnamefont {J.}~\bibnamefont
  {Rajendran}}\ and\ \bibinfo {author} {\bibfnamefont {C.}~\bibnamefont
  {Benjamin}},\ }\bibfield  {title} {\bibinfo {title} {Implementing
  parrondo’s paradox with two-coin quantum walks},\ }\href@noop {} {\bibfield
   {journal} {\bibinfo  {journal} {Royal Society open science}\ }\textbf
  {\bibinfo {volume} {5}},\ \bibinfo {pages} {171599} (\bibinfo {year}
  {2018})}\BibitemShut {NoStop}%
\bibitem [{\citenamefont {Kitagawa}\ \emph {et~al.}(2010)\citenamefont
  {Kitagawa}, \citenamefont {Rudner}, \citenamefont {Berg},\ and\ \citenamefont
  {Demler}}]{Kitagawa2010}%
  \BibitemOpen
  \bibfield  {author} {\bibinfo {author} {\bibfnamefont {T.}~\bibnamefont
  {Kitagawa}}, \bibinfo {author} {\bibfnamefont {M.~S.}\ \bibnamefont
  {Rudner}}, \bibinfo {author} {\bibfnamefont {E.}~\bibnamefont {Berg}},\ and\
  \bibinfo {author} {\bibfnamefont {E.}~\bibnamefont {Demler}},\ }\bibfield
  {title} {\bibinfo {title} {Exploring topological phases with quantum walks},\
  }\href@noop {} {\bibfield  {journal} {\bibinfo  {journal} {Physical Review
  A}\ }\textbf {\bibinfo {volume} {82}},\ \bibinfo {pages} {033429} (\bibinfo
  {year} {2010})}\BibitemShut {NoStop}%
\bibitem [{\citenamefont {Narimatsu}\ \emph {et~al.}(2021)\citenamefont
  {Narimatsu}, \citenamefont {Ohno},\ and\ \citenamefont
  {Wada}}]{narimatsu2021unitary}%
  \BibitemOpen
  \bibfield  {author} {\bibinfo {author} {\bibfnamefont {A.}~\bibnamefont
  {Narimatsu}}, \bibinfo {author} {\bibfnamefont {H.}~\bibnamefont {Ohno}},\
  and\ \bibinfo {author} {\bibfnamefont {K.}~\bibnamefont {Wada}},\ }\bibfield
  {title} {\bibinfo {title} {Unitary equivalence classes of split-step quantum
  walks},\ }\href@noop {} {\bibfield  {journal} {\bibinfo  {journal} {Quantum
  Information Processing}\ }\textbf {\bibinfo {volume} {20}},\ \bibinfo {pages}
  {368} (\bibinfo {year} {2021})}\BibitemShut {NoStop}%
\bibitem [{\citenamefont {Matsuzawa}(2020)}]{matsuzawa2020index}%
  \BibitemOpen
  \bibfield  {author} {\bibinfo {author} {\bibfnamefont {Y.}~\bibnamefont
  {Matsuzawa}},\ }\bibfield  {title} {\bibinfo {title} {An index theorem for
  split-step quantum walks},\ }\href@noop {} {\bibfield  {journal} {\bibinfo
  {journal} {Quantum Information Processing}\ }\textbf {\bibinfo {volume}
  {19}},\ \bibinfo {pages} {1} (\bibinfo {year} {2020})}\BibitemShut {NoStop}%
\bibitem [{\citenamefont {Kadiri}(2023)}]{kadiri2023steered}%
  \BibitemOpen
  \bibfield  {author} {\bibinfo {author} {\bibfnamefont {G.}~\bibnamefont
  {Kadiri}},\ }\bibfield  {title} {\bibinfo {title} {Steered discrete-time
  quantum walks for engineering of quantum states},\ }\href@noop {} {\bibfield
  {journal} {\bibinfo  {journal} {Physical Review A}\ }\textbf {\bibinfo
  {volume} {108}},\ \bibinfo {pages} {012607} (\bibinfo {year}
  {2023})}\BibitemShut {NoStop}%
\bibitem [{\citenamefont {Chandrashekar}\ \emph {et~al.}(2008)\citenamefont
  {Chandrashekar}, \citenamefont {Srikanth},\ and\ \citenamefont
  {Laflamme}}]{chandrashekar2008optimizing}%
  \BibitemOpen
  \bibfield  {author} {\bibinfo {author} {\bibfnamefont {C.~M.}\ \bibnamefont
  {Chandrashekar}}, \bibinfo {author} {\bibfnamefont {R.}~\bibnamefont
  {Srikanth}},\ and\ \bibinfo {author} {\bibfnamefont {R.}~\bibnamefont
  {Laflamme}},\ }\bibfield  {title} {\bibinfo {title} {Optimizing the discrete
  time quantum walk using a su (2) coin},\ }\href@noop {} {\bibfield  {journal}
  {\bibinfo  {journal} {Physical Review A}\ }\textbf {\bibinfo {volume} {77}},\
  \bibinfo {pages} {032326} (\bibinfo {year} {2008})}\BibitemShut {NoStop}%
\bibitem [{\citenamefont {Hoyer}\ and\ \citenamefont
  {Meyer}(2009)}]{hoyer2009faster}%
  \BibitemOpen
  \bibfield  {author} {\bibinfo {author} {\bibfnamefont {S.}~\bibnamefont
  {Hoyer}}\ and\ \bibinfo {author} {\bibfnamefont {D.~A.}\ \bibnamefont
  {Meyer}},\ }\bibfield  {title} {\bibinfo {title} {Faster transport with a
  directed quantum walk},\ }\href@noop {} {\bibfield  {journal} {\bibinfo
  {journal} {Phys. Rev. A}\ }\textbf {\bibinfo {volume} {79}} (\bibinfo {year}
  {2009})}\BibitemShut {NoStop}%
\bibitem [{\citenamefont {Montero}(2013)}]{montero2013unidirectional}%
  \BibitemOpen
  \bibfield  {author} {\bibinfo {author} {\bibfnamefont {M.}~\bibnamefont
  {Montero}},\ }\bibfield  {title} {\bibinfo {title} {Unidirectional quantum
  walks: evolution and exit times},\ }\href@noop {} {\bibfield  {journal}
  {\bibinfo  {journal} {Physical Review A}\ }\textbf {\bibinfo {volume} {88}},\
  \bibinfo {pages} {012333} (\bibinfo {year} {2013})}\BibitemShut {NoStop}%
\bibitem [{\citenamefont {Innocenti}\ \emph {et~al.}(2017)\citenamefont
  {Innocenti}, \citenamefont {Majury}, \citenamefont {Giordani}, \citenamefont
  {Spagnolo}, \citenamefont {Sciarrino}, \citenamefont {Paternostro},\ and\
  \citenamefont {Ferraro}}]{innocenti2017quantum}%
  \BibitemOpen
  \bibfield  {author} {\bibinfo {author} {\bibfnamefont {L.}~\bibnamefont
  {Innocenti}}, \bibinfo {author} {\bibfnamefont {H.}~\bibnamefont {Majury}},
  \bibinfo {author} {\bibfnamefont {T.}~\bibnamefont {Giordani}}, \bibinfo
  {author} {\bibfnamefont {N.}~\bibnamefont {Spagnolo}}, \bibinfo {author}
  {\bibfnamefont {F.}~\bibnamefont {Sciarrino}}, \bibinfo {author}
  {\bibfnamefont {M.}~\bibnamefont {Paternostro}},\ and\ \bibinfo {author}
  {\bibfnamefont {A.}~\bibnamefont {Ferraro}},\ }\bibfield  {title} {\bibinfo
  {title} {Quantum state engineering using one-dimensional discrete-time
  quantum walks},\ }\href@noop {} {\bibfield  {journal} {\bibinfo  {journal}
  {Physical Review A}\ }\textbf {\bibinfo {volume} {96}},\ \bibinfo {pages}
  {062326} (\bibinfo {year} {2017})}\BibitemShut {NoStop}%
\end{thebibliography}%

\end{document}